\documentclass[twocolumn,floats,floatfix,prd,superscriptaddress,nofootinbib,longbibliography]{revtex4-2}

\usepackage{graphicx,epsfig}
\usepackage{amssymb,amsmath,amsthm,amsfonts}
\usepackage{bm}
\usepackage[inline]{enumitem}
\usepackage{tensor}
\usepackage[linktocpage]{hyperref}
\usepackage[caption=false]{subfig}
\usepackage[usenames,dvipsnames]{xcolor}
\usepackage{url}
\usepackage[normalem]{ulem}
\usepackage[inline]{enumitem}
\usepackage{xspace}
\usepackage{comment}
\usepackage{cancel}
\usepackage{multirow}

\def\newacronym#1#2#3{\gdef#1{\gdef#1{#2\xspace}#3 (#2)\xspace}}
\newacronym{\bh}{BH}{black hole}
\newacronym{\bbh}{BBH}{black hole binary}
\newacronym{\gr}{GR}{General Relativity}
\newacronym{\ts}{TS}{topological star}
\newacronym{\qnm}{QNM}{quasinormal mode}

\newcommand{\sapienza}{Dipartimento di Fisica, Sapienza Università 
	di Roma, Piazzale Aldo Moro 5, 00185, Roma, Italy}
\newcommand{\infn}{INFN, Sezione di Roma, Piazzale Aldo Moro 2, 00185, Roma, Italy}

\newcommand{\centra}{CENTRA, Departamento de Física, Instituto Superior Técnico – IST\\
Universidade de Lisboa – UL, Avenida Rovisco Pais 1, 1049-001 Lisboa, Portugal}

\begin{document}

\title{Extreme mass ratio inspirals around topological stars
}

\author{Marco Melis}
\email{marco.melis@uniroma1.it}
\affiliation{\sapienza}
\affiliation{\infn}

\author{Richard Brito}
\email{richard.brito@tecnico.ulisboa.pt}
\affiliation{\centra}

\author{Paolo Pani}
\email{paolo.pani@uniroma1.it}
\affiliation{\sapienza}
\affiliation{\infn}

\begin{abstract}
  We study a point scalar charge in circular orbit around a topological star, a regular, horizonless soliton emerging from dimensional compactification of Einstein-Maxwell theory in five dimensions, which could describe qualitative properties of microstate geometries for astrophysical black holes.
  This is the first step towards studying extreme mass-ratio inspirals around these objects.
  We show that when the particle probes the spacetime close to the object, the scalar-wave flux deviates significantly from the corresponding black hole case. Furthermore, as the topological star approaches the black-hole limit, the inspiral can resonantly excite its long-lived modes, resulting in sharp features in the emitted flux.
  Although such resonances are too narrow to produce detectable dephasing, we estimate that a year-long inspiral down to the innermost stable circular orbit could accumulate a significant dephasing for most configurations relative to the black hole case. While a full parameter-estimation analysis is needed, the generically large deviations are likely to be within the sensitivity reach of future space-based gravitational-wave detectors.
\end{abstract}

\maketitle

\section{Introduction}
The information loss paradox~\cite{Hawking:1976ra,Polchinski:2016hrw}, the presence of curvature singularities, and other fundamental issues associated with black hole~(BH) horizons have long motivated the search for alternative models of compact objects. These BH mimickers --~horizonless and regular objects with properties similar to classical BHs~-- offer a compelling avenue to address such conundrums~\cite{Cardoso:2019rvt,Buoninfante:2024oxl,Carballo-Rubio:2025fnc}.
Among various proposals, one of the few rooted in a fundamental theory is the fuzzball paradigm of string theory, which interprets a classical BH as a thermodynamic approximation of an enormous number of regular quantum states~\cite{Mathur:2009hf,Bena:2022rna,Bena:2022ldq}.
In the classical limit, the latter are represented by ``microstate geometries'' --~solitonic solutions that share the same mass and charges as a BH but exhibit a fundamentally different internal structure, where the horizon is replaced by a smooth, horizonless cap~\cite{Bena:2006kb,Bena:2016ypk,Bena:2017xbt,Bah:2021owp,Bah:2022yji}.

The fuzzball construction relies on two key aspects: the inherently higher-dimensional nature of microstate solutions and their complex topologies, which stabilize the horizon-scale structure and prevent its collapse.
These solutions are particularly involved and, from the four-dimensional perspective, do not have any spatial symmetries. This has so far prevented dynamical studies, (see~\cite{Ikeda:2021uvc} for an exception using a test field), which are instead necessary to predict the gravitational-wave~(GW) signatures of this model and distinguish it to the classical BH prediction. 

A notably simple model that integrates key fuzzball concepts was introduced some years ago in~\cite{Bah:2020ogh}, revealing singularity-free solitons within five-dimensional Einstein-Maxwell theory. These solutions, termed topological stars~(TSs), possess topological cycles sustained by magnetic flux. TSs exhibit several intriguing properties, including their four-dimensional reduction --~obtained by compactifying the fifth dimension on a circle~-- which asymptotically resembles magnetic BHs but lacks an event horizon.
Although these solutions have electromagnetic charges and therefore are not a good proxy for neutral BHs\footnote{Extensions of these solutions can be constructed such that they are globally neutral and asymptote to a Schwarzschild BH~\cite{Bah:2022yji}, although at the expenses of introducing more complexity and breaking spherical symmetry.}, they could capture  qualitative properties of microstate geometries for astrophysical BHs.

These features have motivated recent studies exploring ray tracing~\cite{Heidmann:2022ehn}, the dynamics of test fields in these geometries~\cite{Heidmann:2023ojf,Bianchi:2023sfs}, and their linear response~\cite{Dima:2024cok,Bena:2024hoh,Bianchi:2024vmi,Cipriani:2024ygw,Dima:2025zot}.
In particular, Refs.~\cite{Dima:2024cok,Bena:2024hoh,Dima:2025zot} studied the linear dynamics of TSs, to compute their quasinormal modes~(QNMs) and ringdown waveforms. These studies found strong numerical evidence that TSs are linearly stable in the most interesting region of the parameter space.\footnote{There exists a radial unstable mode with momentum along the extra dimension, which is directly connected to the Gregory-Laflamme instability~\cite{Gregory:1993vy} of charged black strings~\cite{Miyamoto:2006nd,Stotyn:2011tv,Bah:2021irr}. However, this mode only affects TSs with small compactness, leaving the most interesting region of the parameter space available.}

Recently, Refs.~\cite{Bianchi:2024rod,DiRusso:2025lip} studied the scalar emission by a point test scalar charge moving around a TS.
The scope of this paper is to further study the dynamics and emission of a point test charge in orbital motion around these geometries, as a proxy for extreme mass-ratio inspirals~(EMRIs) that can be used to test the absence of BH horizons.

EMRIs are among the most interesting sources for future space missions such as LISA~\cite{LISA:2024hlh}, and potentially also for ground-based detectors, should subsolar compact objects exist~\cite{Barsanti:2021ydd}.
An EMRI is typically composed by a stellar-mass compact object (the secondary) orbiting around a much heavier one (the primary), with a mass ratio $\mu/M\ll1$, allowing for perturbative analysis~\cite{Poisson:2011nh,Barack:2018yvs}.
To leading order, the secondary follows geodesics of the primary background spacetime; the geodesic parameters evolve adiabatically due to radiation-reaction effects driven by the emission of GWs at infinity or into the horizon of the primary, if the latter is a BH.

EMRIs are unique probes of the strong gravity regime, as they can perform $\approx 10^5$ orbits in LISA band before plunging, spending most of the time in the vicinity of the primary at relativistic orbital velocity. The corresponding long GW signal carries extremely accurate information about both the spacetime structure around the primary~\cite{Kesden:2004qx,Macedo:2013qea,Babak:2017tow,Pani:2019cyc,Piovano:2022ojl,Bianchi:2020bxa,Destounis:2023khj,Ghosh:2024arw}, the dynamics of the underlying theory of gravity~\cite{Pani:2011xj,Cardoso:2011xi,Yunes:2011aa,Cardoso:2018zhm,Maselli:2021men,Barsanti:2022vvl,Speri:2024qak} (see \cite{Barausse:2020rsu,LISA:2022kgy,Cardenas-Avendano:2024mqp} for some reviews), the environment around the primary~\cite{Barausse:2014tra,Hannuksela:2018izj,DellaRocca:2024sda,Cardoso:2022whc,Dyson:2025dlj}, and the physics at the horizon scale~\cite{Pani:2010em,Cardoso:2019nis,Datta:2019epe,Maggio:2021uge,Datta:2024vll}.

Studies of EMRIs around horizonless compact objects to constrain horizon-scale physics were so far limited to phenomenological models (with the exception of EMRIs around gravastars~\cite{Mazur:2004fk,Mazur:2001fv} studied in~\cite{Pani:2010em}) wherein the interior of the object is modelled by an effective reflectivity function~\cite{Datta:2019epe,Maggio:2021uge,Datta:2024vll}. 
In this paper, building on~\cite{Bianchi:2024rod}, we extend these studies to the case of TSs.
At variance with the analysis in~\cite{Bianchi:2024rod}, we solve the full numerical problem for generic circular orbits, also beyond the weak-field approximation. Furthermore, we compare the result with the BH case and unveil the fact that, in the strong-field regime, it is possible to excite the long-lived QNMs of a TS~\cite{Bena:2024hoh,
Dima:2024cok,Dima:2025zot}. This yields resonances in the energy flux emitted at infinity that are absent in the BH case. Finally, we also estimate the difference between TSs and BHs in terms of scalar wave dephasing, as a first step to quantify detectability.

\section{Five-dimensional theory and background solutions}

Let us consider the Einstein-Maxwell theory in 5D
\begin{equation}
    S_{EM} = \int d^5x \sqrt{-g} \left( \frac{1}{2 \kappa_5^2} R -\frac14 F_{\mu\nu}F^{\mu\nu} \right)\,,
\end{equation}
where we will set units such that $\kappa_5^2$ has dimensions of a length.
From the above action, one can derive the following field equations
\begin{align}
   & R_{\mu\nu} - \frac12 g_{\mu\nu} R + \kappa_5^2 \left( 
F_{\mu\rho}F^\rho{}_\nu + \frac14 g_{\mu\nu}F_{\rho\sigma}F^{\rho\sigma} \right) = 0 \,,
\\
& \nabla^\nu F_{\mu\nu} = 0 \,.
\end{align}
The theory admits two solutions~\cite{Bah:2020ogh}, both carrying a magnetic charge $P$ and described by the following line element 
\begin{align}
    & ds^2 = -f_S dt^2 + f_B dy^2 + \frac{dr^2}{f_B f_S} + r^2 d\Omega_2^2 \,,
    \\
    & F = P \sin \theta d\theta \wedge d\phi \,,
\end{align}
with 
\begin{equation}
    f_S = 1- \frac{r_S}{r}\,, \hspace{0.3cm} f_B = 1- \frac{r_B}{r}\,, \hspace{0.3cm} P = \pm \frac{1}{\kappa_5}\sqrt{\frac{3 r_S r_B}{2}} \,.
\end{equation}
One solution is a magnetized black string ($r_S \geq r_B$) while the other is a regular soliton known as TS ($r_B > r_S$).
The fifth dimension is compact and described by the coordinate $y$, with period $2 \pi R_y$, where $R_y$ is its radius\footnote{The radius $R_y$ of the $y$-circle and the parameters $r_B$ and $r_S$ are related by an orbifold condition that also implies an upper bound $2 r_B \leq n R_y$, where $n$ is an integer~\cite{Bah:2020ogh}. 
See~\cite{Bah:2020pdz,Bah:2021rki} for more details about this relation.}.
Asymptotically the two solutions are given by four-dimensional Minkowski times a circle parametrized by $y$. After performing a dimensional reduction, the ADM mass in four dimensions is 
\begin{equation}\label{eq:4DADMmass}
    M = \frac{2 \pi}{\kappa_4^2}(2 r_S + r_B) \,.
\end{equation}
where we introduced the coupling $\kappa_4^2 \equiv \kappa_5^2 / (2 \pi R_y)$, and the magnetized black string becomes a magnetized BH in four dimensions.\footnote{ By using an electromagnetic duality~\cite{Bah:2020pdz}, it is equivalently possible to describe electrically charged BHs and TSs in this theory.} 

Depending on its compactness and its photon spheres (as discussed in Sec.~\ref{sec:geodesics}), the TS can be distinguished in \textit{first kind} ($3/2 \leq r_B / r_S \leq 2$) and \textit{second kind} ($1 \leq r_B / r_S \leq 3/2$).
Note that when $r_B\to0$ the solution reduces to the ordinary Schwarzschild BH, while for $r_B\to r_S$ the TS approaches the metric of an extremal BH.
In five dimensions TSs are everywhere regular and were found to be linearly stable, except for the well know Gregory-Laflamme instability that affects black strings for $r_S > 2 r_B$. Indeed, by performing a double Wick rotation
$(t, y, r_S , r_B) \to (iy, it, r_B , r_S)$ 
this instability also affects TSs with $r_B > 2 r_S$~\cite{Bah:2021irr}. The most interesting case of compact TSs ($r_S<r_B<2 r_S$) that can approach the BH limit was found to be linearly stable against gravito-electromagnetic-scalar perturbations~\cite{Bena:2024hoh,
Dima:2024cok,Dima:2025zot}.

For convenience, in the following we will refer to a magnetized black string ($r_S \geq r_B$) simply as magnetized BH, having in mind that the extra dimension is small compared to the size of the compact objects.

\section{Geodesic equations:\\
null and time-like circular orbits}
\label{sec:geodesics}
Let us study the geodetic motion of null and time-like particles in the background of the TS (or the magnetized BH). 
For simplicity we restrict the analysis only to circular geodesics on the equatorial plane, i.e. $\dot{\theta} = 0$ and $\theta = \pi / 2$. The Lagrangian for a free particles is given by 
\begin{equation}
    \mathcal{L} = \frac12 \left[ -f_s \dot{t}^2 + f_b \dot{y}^2 + \frac{\dot{r}^2}{f_B f_S} +r^2 \dot{\varphi}^2 \right]\,,
\end{equation}
where the dot denotes the derivative with respect to an affine parameter $\lambda$. The $5-$momentum $P_\mu = \partial \mathcal{L} / \partial \dot{x}^\mu$ of the test particle reads
\begin{align}
     P_t &= - f_s \dot{t} = -E \\
     P_y &= f_b \dot{y} = p \\
     P_r &= \frac{\dot{r}}{f_b f_s} \\
     P_\varphi &= r^2 \dot{\varphi} = L \,,
\end{align}
where $E$, $L$, and $p$ are integrals of motion associated to isometries along the $t$, $\varphi$, and $y$ directions, respectively.
The Hamiltonian of the particle is given by 
\begin{equation}
    H = \frac12 g_{\mu\nu} P^\mu P^\nu = -\frac{\mu^2}{2} \,,
\end{equation}
where $\mu$ is the mass of the free particle that for null particles is zero.
If we neglect the momentum $p$ along the compact dimension, the radial geodesic equations is the following
\begin{equation}
    \dot{r}^2 = E^2 - V_{\rm eff} \,,
\end{equation}
where the effective potential $V_{\rm eff}$ is 
\begin{equation}
    V_{\rm eff} = \frac{r_B}{r}E^2 + f_S f_B \left( \mu^2 +\frac{L^2}{r^2} \right) \,. 
\end{equation}
Circular geodesics are defined by vanishing radial velocity and acceleration, $\dot{r} = \Ddot{r} = 0$, and can be parametrized in terms of the energy $E$ and of the angular momentum $L$, given by
\begin{equation}\label{eq:energy_angmom}
    \frac{E}{\mu} = \sqrt{\frac{2 (r-r_S)^2}{r(2r-3r_S)}} \,, \hspace{0.7cm}
    \frac{L}{\mu} = \sqrt{\frac{r_S \, r^2}{2r-3r_S}} \,.
\end{equation}
The roots of $\dot{r} = \Ddot{r} = 0$ are
\begin{align}
    \mu &= 0 \,: \hspace{0.6cm} r_1 = r_B \,, \hspace{0.4cm} r_2 = \frac32 r_S \\
    \mu &\neq 0 \,: \hspace{0.6cm} r_1 = r_B \,, \hspace{0.4cm} r_{2\pm} =L^2 \frac{ 1\pm \sqrt{L^2 - 3 r_S^2 \mu^2}}{r_S \mu^2}\,. 
\end{align}
The case $\mu = 0$ corresponds to null circular orbits, in particular for \textit{first kind} TSs there is a single unstable photon sphere at $r_2$, while \textit{second kind} TSs have an unstable photon sphere at $r_2$ and a stable one at $r_1$.
The case $\mu \neq 0$ describes circular orbits for massive probes. When the two roots coincide, $r_{2+} = r_{2-}$, we find the innermost stable circular orbit (ISCO)
\begin{equation}
    r_{\rm ISCO} = 3 r_S \,.
\end{equation}
The $t$-component of the four-velocity and the angular velocity are given by
\begin{equation}
    u^t = \frac{dt}{d\tau} = \frac{E}{\mu f_S} \,, \hspace{0.7cm} \Omega = \frac{d \varphi}{dt} = u^t \frac{L}{\mu} \,, 
\end{equation}
that in the case of circular orbits with $r=r_0$ are
\begin{equation}\label{eq:Keplerian_velocity}
    u^t = \frac{1}{\sqrt{1-\frac{r_S}{r_0} - r_0^2 \Omega^2}} \,, \hspace{0.7cm} \Omega = \sqrt{\frac{r_S}{2 r_0^3}} \,.
\end{equation}
The Schwarzschild limit is recovered by setting $r_B = 0$, in which case $r_S$ is the radial position of the Schwarzschild horizon.\footnote{The Keplerian orbital velocity in Eq.~\eqref{eq:Keplerian_velocity} depends on the gravitational mass in 5D, i.e. $r_S/2$, rather than on the 4D ADM mass $M$ defined by Eq.~\eqref{eq:4DADMmass}. This distinction between the 5D and 4D masses is a peculiarity of the TS metric. 
In what follows, we will normalize all dimensionful quantities using the 5D mass, as it provides a more natural reference scale within the 5D framework.}

For later purposes, it is useful to compute the time radiation in radial infall takes to reach the boundary of the TS. Setting $\mu=0=L$ in the above equations, we can compute
\begin{equation}
    \Delta T =\int_{r_i}^{r_B}dr \frac{\dot t}{\dot r}\,,
\end{equation}
where $r_i$ is the initial radial distance. The integral above has a closed form expression for any $r_B$, but it is particularly simple in the BH limit, $r_B\to r_S$:
\begin{equation} \label{trapping}
    \Delta T\sim \frac{\pi r_S^{3/2}}{\sqrt{r_B-r_S}}\,.
\end{equation}
As expected this quantity diverges as $r_B\to r_S$, namely as the TS approaches the extremal BH solution. 

\section{Scalar field equation, energy flux, and orbital evolution}
Consider a test scalar charge orbiting in the background of the TS (or the magnetized BH) along a circular orbit with null momentum along the $y-$direction ($P_y = 0=p$).  Due to the spherical symmetry of the background we restrict the geodesic to be a planar circular orbit ($\theta = \pi /2 $) with radial distance $r_0\geq r_{\rm ISCO}$. 

The sourced scalar field $\Phi$ satisfies the following inhomogeneous Klein-Gordon equation
\begin{equation}
    \Box \Phi = \rho_p \,,
\end{equation}
where $\rho_p$ is the scalar charge density defined by the integral over the affine parameter $\tau$ along the worldline of the particle 
\begin{equation}
    \rho_p = \mu \, q \int d\tau \frac{1}{\sqrt{-g}} \delta^{(5)}(x^\rho - z^\rho(\tau)) \,.
\end{equation} 
If $\Phi$ is taken to be dimensionless, then $q$ has dimensions of a length.
Since the particle follows the time-like geodesic defined by circular orbits at $r_0$, the five-dimensional Dirac delta function can be decomposed in the following way 
\begin{align}
    \delta^{(5)}(x^\rho - z^\rho(\tau))  = &\delta(t-u^t \tau) \delta(r-r_0) \delta(y)\, \nonumber\\ 
    &\delta \left(\theta - \frac{\pi}{2}\right) \delta (\varphi - \Omega t) \,.
\end{align}
Using the fact that $\sqrt{-g} = r^2 \sin \theta$ and the closure relation of spherical harmonics, after performing the integral we get 
\begin{align}
    \rho_p = \frac{\mu q}{u^t r^2} &\delta(r-r_0) \sum_n \frac{e^{i n \frac{y}{R_y}}}{2 \pi R_y} \nonumber \\ 
    &\times \sum_{lm} Y_{lm}(\theta,\varphi) Y_{lm}^*\left(\frac{\pi}{2},0\right) e^{-im\Omega t} \,,
\end{align}
where we have used the relation 
\begin{equation}
    \delta(y) = \sum_n \frac{e^{i n \frac{y}{R_y}}}{2 \pi R_y} \,.
\end{equation}
By making use of the Fourier transform and of the identity 
\begin{equation}
    \delta(\omega - m \Omega) = \int \frac{dt}{2\pi} e^{i 
(\omega - m\Omega) t} \,,
\end{equation}
the source $\rho_p$ can be written in the following more convenient form
\begin{align}
    \rho_p = \sum_{nlm} \int \frac{d\omega}{2\pi} e^{i\omega t} \tilde{S}_{mln}(\omega,r) e^{in \frac{y}{R_y}} Y_{lm}(\theta,\varphi) \,,
\end{align}
where we defined
\begin{equation}
    \tilde{S}_{mln}(\omega,r) \equiv \frac{\mu q}{u^t r^2 R_y} \delta(\omega-m\Omega)\delta(r-r_0)Y_{lm}^*\left( 
\frac{\pi}{2},0\right) \,.
\end{equation}
If we expand the scalar field in spherical harmonics and use the Fourier transform
\begin{equation}\label{eq:expansion_scal}
    \Phi = \sum_{nlm} \int \frac{d\omega}{2\pi} e^{-i \omega t} e^{i n \frac{y}{R_y}} \psi_{nlm}(r) Y_{lm}(\theta,\varphi) \,,
\end{equation}
we can isolate the radial part of the Klein-Gordon equation to get 
\begin{align} \label{eq:KG_eq_radial}
    &\left[ f_S f_B \partial_r^2 + \frac{f_S + f_B}{r}\partial_r + \frac{\omega^2}{f_S} - \frac{n^2}{f_B R_y^2} - \frac{l(l+1)}{r^2} \right] \psi_{nlm}(r)\nonumber \\
    &= \tilde{S}_{nlm} (\omega,r) \,.
\end{align}
By introducing the tortoise coordinate $d\rho/dr = f_S^{-1}f_B^{-1/2}$ and performing the following field redefinition
\begin{equation}
    \Psi_{nlm} = r f_B^{1/4} \psi_{nlm}\,,
\end{equation}
we get a single Schrödinger-like equation,
\begin{align}\label{eq:Schro_eq}
    \left[\partial_\rho^2 + (\omega^2 -V_{\rm eff}) \right] \Psi_{nlm} = \mathcal{S}_{nlm}(\omega,\rho)\,,
\end{align}
where the effective potential is given by
\begin{align}
    V_{\rm eff} = \frac{r-r_S}{r-r_B}&\biggl\{ \frac{n^2}{R_y^2} +\frac{1}{16 r^5} \biggl[ 16 r^3 l(l+1) \nonumber \\ 
    &+21 r_B^2 r_S +8 r^2 (r_B -2 l r_B -2 l^2 r_B +2 r_S) \nonumber \\
    &- 9 r r_B (r_B +4 r_S) \biggr] \biggr\}\,,
\end{align}
and the source is 
\begin{equation}\label{eq:source_eq}
    \mathcal{S}_{nlm}(\omega,\rho) \equiv \frac{\mu q}{u^t R_y} \frac{f_S f_B^{1/4}}{r} \delta(\omega - m\Omega) \delta(r-r_0) Y_{lm}^*\biggl(\frac{\pi}{2},0\biggr)
\end{equation}
From now on we will drop the $nlm$ indexes for notation convenience. 
Furthermore, from Eq.~\eqref{eq:KG_eq_radial} one can see that perturbations have a mass proportional to $n/R_y$ and only the $n=0$ mode is massless and propagates to infinity as a free wave.
In the phenomenologically interesting limit, $R_y\ll r_S$, the $n\neq0$ modes have a large mass and are therefore exponentially suppressed on a length scale  $R_y\ll r_S$. Thus, from now on we will focus on the massless $n=0$ modes only.

Following the usual procedure that makes use of the Green function solution of the equation sourced by a Dirac delta source $\delta(\rho -\rho')$, we can write the general solution as
\begin{equation}
    \Psi(\rho) = \frac{\Psi_+}{W} \int_{\rho_{\rm in}}^\rho \mathcal{S}(\omega,\rho') \Psi_- d \rho' + \frac{\Psi_-}{W}\int_\rho^{+\infty} \mathcal{S}(\omega,\rho') \Psi_+ d\rho' \,,
\end{equation}
where $\rho_{\rm in}$ is the inner-boundary value of the tortoise coordinate, i.e. $\rho_{\rm in}=\{-\infty,\rho(r_B)\}$ for the magnetized BH and TS, respectively, whereas $\Psi_-$ and $\Psi_+$ are two independent solutions of the homogeneous equation, satisfying the boundary condition presented below. Their corresponding Wronskian reads
\begin{equation}
    W = \Psi_- \frac{d \Psi_+}{d r} - \Psi_+ \frac{d \Psi_-}{d r}  \,,
\end{equation}
and does not depend on $r$ by construction, as can be easily seen by means of Eq.~\eqref{eq:Schro_eq} without the source term.
While asymptotically at infinity ($r \rightarrow +\infty$) these two solutions behave as 
\begin{align}\label{eq:sol_inf_BC}
    \Psi_- &\sim A_{\rm out} e^{i \omega \rho} + A_{\rm in} e^{-i \omega \rho} \\
    \Psi_+ &\sim e^{i \omega \rho}  \,,
\end{align}
their asymptotic behavior near the inner boundary depends on the background spacetime \cite{Dima:2024cok,Dima:2025zot}.

For a magnetized BH, the boundary condition as $r \rightarrow r_S $ is
\begin{align}\label{eq:BC_rs_BH}
    \Psi_- &\sim (r-r_S)^{-i k_S \omega} \\
    \Psi_+ &\sim B_{\rm out} (r-r_S)^{i k_S \omega} + B_{\rm in} (r-r_S)^{-i k_S \omega}\,,
\end{align}
where $k_S \equiv r_S^{3/2} / \sqrt{r_S - r_B}$, and we used the radial coordinate $r$ for convenience.
In the case of a TS, the boundary condition at $r \rightarrow r_B$ reads 
\begin{align}\label{eq:BC_rb_TS}
   \Psi_- &\sim (r-r_B)^{1/4} \\
   \Psi_+ &\sim C \, (r-r_B)^{1/4} + D (r-r_B)^{1/4} \log(r-r_B) \,.
\end{align}

Using $dW/dr=0$, we can use the behavior of the two independent solutions at infinity to get $W = 2 i \omega A_{\rm in}$. 
The final solution at asymptotic infinity is given by 
\begin{align}\label{eq:scal_sol_inf}
    &\Psi_\infty = \frac{e^{i \omega \rho}}{2 i \omega A_{\rm in}} \int_{\rho_{\rm in}}^{+\infty} \mathcal{S}(\omega,\rho') \Psi_-(\rho') d \rho' \nonumber \\
    &= \frac{e^{i\omega \rho}}{2 i \omega A_{\rm in}} \int_{r_{\rm in}}^{+\infty} \frac{dr'}{f_S f_B^{1/2}} \mathcal{S}(\omega,r') \Psi_-(r')  \nonumber \\
    &= \frac{e^{i\omega \rho}}{2 i \omega A_{\rm in}} \frac{\mu q}{u^t|_{r_0} R_y} \frac{\Psi
_-(r_0)}{r_0 f_B^{1/4}(r_0)} \delta(\omega - m \Omega) Y^*_{lm}\left( \frac{\pi}{2},0 \right) \,.
\end{align}
where $r_{\rm in} = \{r_S, r_B\}$ for magnetized BH or TS, respectively.
Similarly, the final solution near the horizon of the magnetized BH is 
\begin{align}\label{eq:scal_sol_hor}
     \Psi_{r_S} &= \frac{e^{-i \omega \rho}}{2 i \omega A_{\rm in}} \int_{-\infty}^{+\infty} \mathcal{S}(\omega,\rho')\Psi_+(\rho')d\rho' \nonumber \\
    &= \frac{e^{-i \omega \rho}}{2 i \omega A_{\rm in}} \frac{\mu q}{u^t|_{r_0} R_y} \frac{\Psi_+(r_0)}{r_0 f_B^{1/4}(r_0)} \delta(\omega - m\Omega) Y_{lm}^* \biggl( \frac{\pi}{2},0 \biggr) \,.
\end{align}
The scalar energy flux at infinity and at $r_S$ is defined by 
\begin{equation}
    \dot{E}_{\infty,r_S} = \lim_{r \rightarrow +\infty , r_S} \int d\theta d\varphi \sqrt{-g} T^r_t \,,
\end{equation}
where the stress-energy tensor of the scalar field is 
\begin{equation}
    T_{\mu\nu} = \partial_\mu \Phi \partial_\nu \Phi - \frac12 g_{\mu\nu} \partial_\rho \Phi \partial^\rho \Phi \,.
\end{equation}
Finally, if we consider the decomposition~\eqref{eq:expansion_scal} with $n=0$ and using Eq.~\eqref{eq:scal_sol_inf} we get 
\begin{equation}\label{eq:flux_inf}
    \Dot{E}_{\infty,r_S} = \sum_{lm} (m \Omega)^2 |\Psi_{\infty,r_S}|^2 \equiv \sum_{lm}\Dot{E}^{lm}_{\infty,r_S}\,.
\end{equation}

\subsection{Effective trapping of radiation}
Clearly, the energy flux at the horizon, $\dot E_{r_S}$, is present only in the BH case. However, even in the absence of a horizon, efficient trapping of radiation within the object could effectively produce an effect similar to the energy loss at the horizon~\cite{Maselli:2017cmm,Datta:2019epe,Maggio:2021uge}.
This occurs if the trapping time of radiation near the object, $T_{\rm trap}$, exceeds the typical radiation-reaction time scale of the binary~\cite{Datta:2019epe}. In the Newtonian limit (taking the gravitational case as a proxy\footnote{In addition of being more interesting from a phenomenological perspective, the gravitational case is also more conservative since, as discussed later, the gravitational flux is significantly higher than the scalar one, resulting in a \emph{shorter} $T_{\rm RR}$.}), the latter reads
\begin{equation}
    T_{\rm RR}\sim \frac{5}{64}\left(\frac{r_0}{M}\right)^4\frac{M^2}{\mu}\,,
\end{equation}
where $M$ is given in Eq.~\eqref{eq:4DADMmass} and we shall use units such as $\kappa_4^2={8\pi}$.
We can estimate the trapping time using Eq.~\eqref{trapping}. The most relevant regime is when $r_B\to r_S$, since the trapping time diverges,
\begin{equation}
    T_{\rm trap}\sim \frac{\pi r_S^{3/2}}{\sqrt{r_B-r_S}}\,.
\end{equation}
By comparing $T_{\rm trap}$ with $T_{\rm RR}$, we find that trapping is inefficient when
\begin{equation} \label{trappingcond}
    \frac{r_B}{r_S}-1\gg \frac{4096\pi^2}{25}\frac{r_S^2}{M^2}\left(\frac{M}{r_0}\right)^8\frac{\mu^2}{M^2}\approx 10^{-15}\left(\frac{\mu}{10^{-6}M}\right)^2\,,
\end{equation}
where in the last estimate we considered $r_S\approx M$ and $r_0\approx 6M$. In other words, unless $r_B$ is close to $r_S$ within one part in $10^{15}$, effective trapping can be neglected and the only source of energy loss for a TS is the flux at infinity.

\section{Orbital evolution}
EMRIs are typically studied under the assumption that the system evolves adiabatically under radiation reaction. 
This is true as long as the radiation reaction time scale is much longer than the orbital period of the secondary around the primary compact object. 
Under this assumption, to leading order the evolution of the parameters describing the system is governed by the balance equation
\begin{equation}\label{eq:balance_eq}
    \Dot{E}_b = \Dot{r} \frac{dE_b}{dr} = - \Dot{E}^s_{\rm tot} \,,
\end{equation}
where $E_b$ is the binding energy, given by Eq.~\eqref{eq:energy_angmom}, and $\Dot{E}^s_{\rm tot}$ is the total emitted scalar energy flux
\begin{align}
    \Dot{E}^s_{\rm tot} &= \Dot{E}^s_\infty \hspace{2cm} \text{for the TS,} \nonumber \\
    \Dot{E}^s_{\rm tot} &= \Dot{E}^s_{r_S} +\Dot{E}^s_\infty
    \hspace{1cm} \text{for the BH,}
\end{align}
where for the TS we assumed that effective trapping is negligible, i.e. Eq.~\eqref{trappingcond}.
The above balance equation can be used to estimate the time necessary for the test scalar charge to reach the ISCO 
\begin{equation}
    t_{\rm ISCO} = \int_{r_{\rm ini}}^{r_{\rm ISCO}} dr \biggl( -\frac{dE_b}{dr} \biggr) \frac{1}{\Dot{E}^s_{\rm tot}} \,.
\end{equation}

The phase of the scalar wave emitted by the test charge can be computed as
\begin{equation}\label{eq:phase_eq}
    \phi_s = \int_{r_{\rm ini}}^{r_{\rm fin}} dr \, \frac{\Omega(r)}{\Dot{r}} \,.
\end{equation}
\section{Analytical solution at large orbital distance}\label{sec:analytical_solution}
Before solving Eq.~\eqref{eq:Schro_eq} numerically, let us provide an analytical solution valid in the low-frequency regime, which allows us to compute the energy flux for large orbital separations (see \cite{PhysRevD.47.1497,Poisson:1994yf,Kanti:2002nr,Chen:2007ay,Creek:2007pw,Brito:2012gw} for similar analyses in other contexts). We follow the standard matching asymptotics technique: namely, we match the solution near the inner boundary to the one valid asymptotically at infinity; such matching is possible only in the low-frequency approximation, $r_S\Omega \ll 1$. This allows us to compute the amplitude of the solution at $r_0 \gg r_S$ and the corresponding scalar flux analytically.

\subsection{Magnetized BH}
If we define $\Delta \equiv r^2 f_B f_S$, the homogeneous version of Eq.~\eqref{eq:KG_eq_radial} with $n=0$ can be written in the following convenient way
\begin{equation}\label{eq:KG_rad_2}
    \frac{d}{dr} \left( \Delta \frac{d \psi}{dr} \right) + \frac{r^3 \omega^2 -l(l+1)(r-r_S)}{r-r_S}\psi = 0 \,.
\end{equation}
Let us first focus on the magnetized BH.
Consider the following change of variable of the radial coordinate $r$, introducing a new coordinate $h$ that approaches zero at the horizon,
\begin{equation}
    \hspace{0.3cm} h \equiv \frac{r-r_S}{r-r_B}\,.
\end{equation}
In the $h \ll 1$ limit, Eq.~\eqref{eq:KG_rad_2} can be written as
\begin{align}
    &h(1-h) \frac{d^2 \psi}{dh^2} + (1-h)\frac{d \psi}{dh} \nonumber \\
    &+\biggl[ \frac{r_S^3 \omega^2}{(r_S-r_B)h(1-h)} - \frac{l(l+1)}{1-h} \biggr]\psi = 0 \,.
\end{align}
By making use of the field redefinition $\psi(h) \equiv h^\alpha (1-h)^\beta F(h)$, the above equation can be recast as an hypergeometric differential equation
\begin{align}
    h(1-h)\frac{d^2 F}{dh^2} +[c-(a+b+1)h]\frac{dF}{dh} - ab F = 0 \,,\label{eq:hypergeom}
\end{align}
where $a = b = \alpha + \beta$, $c = 1+2 \alpha$,
whereas $\alpha$ and $\beta$ satisfy the following algebraic equations
\begin{align}
    \alpha^2 + \frac{r_S^3 \omega^2}{r_S - r_B} &= 0 \\
    \beta^2 - \beta - \alpha^2 -l(l+1) &= 0
\end{align}
with solutions 
\begin{align}
    \alpha_{\pm} &= \pm i \sqrt{\frac{r_S^3}{r_S-r_B}} \, \omega = \pm i k_S \, \omega \\
    \beta_{\pm} &= \frac12 [1 \pm \sqrt{(1+2l)^2 + 4\alpha^2}] \,.
\end{align}
From the general solution of the hypergeometric differential equation we get 
\begin{align}
    &\psi(h) = A_1 h^\alpha (1-h)^\beta F(a,b,c;h) \nonumber\\
    &+ B_1 h^{-\alpha} (1-h)^\beta F(a-c+1,b-c+1,2-c;h) \,,
\end{align}
where $F(a,b,c;h)$ is the hypergeometric function. We require the hypergeometric functions to be convergent for $|h|\leq 1$, i.e. to have ${\rm Re}[c-a-b] > 0$, thus selecting the solutions $\beta = \beta_-$ and $\alpha = \alpha_-$. If we expand the solution near the horizon $r_S$ we get 
\begin{equation}\label{eq:hyper_sol_BC}
    \psi(h) = A_1 h^{-i k_S \omega} + B_1 h^{i k_S \omega} \,,
\end{equation}
and imposing only ingoing waves we require $B_1 = 0$. We want to match the near horizon with the far-field solution in an intermediate region which exists in the low-frequency regime. For this purpose, using the properties of the hypergeometric function \cite{abramowitz+stegun}, we change the argument of the hypergeometric function from $h$ to $1-h$ to get 
\begin{align}
    \psi(h) = &A_1 h^\alpha (1-h)^\beta \times \nonumber \\
    &\biggl[ 
\frac{\Gamma(1+2\alpha) \Gamma(1-2\beta)}{\Gamma(1+\alpha-\beta)^2} F(a,b,a+b-c+1;1-h) \nonumber \\
 &+(1-h)^{1-2\beta} \frac{\Gamma(1+2\alpha)\Gamma(2\beta -1)}{\Gamma(\alpha+\beta)^2} \times \nonumber \\
 &F(c-a,c-b,c-a-b+1;1-h) \biggr]\,.
\end{align}

Since we are interested in the low-frequency approximation, i.e. $r_S \, \Omega = r_S \, \omega / m \ll 1$, as long as the $\omega r_S \ll 1$ condition is fulfilled, $\beta \leq 0$. Expanding the solutions for $1-h \ll 1$, namely $r \gg r_S$, we get
\begin{align}
    \Psi(r) &= r f_B^{1/4} \psi(r) \nonumber \\
    &\sim \frac{r^{l+1}}{r_S^{3/4}(r_S-r_B)^{l+1/4}}\frac{\Gamma(1+2\alpha)\Gamma(1-2\beta)}{\Gamma(1+\alpha-\beta)^2} \,,
\end{align}
where we obtained $A_1 = r_S^{-3/4}(r_S-r_B)^{-1/4}$ by comparing Eq.~\eqref{eq:hyper_sol_BC} with the boundary condition in Eq.~\eqref{eq:BC_rs_BH}.

Let us now study the far-field limit solution of Eq.~\eqref{eq:Schro_eq}, following the procedure outlined in \cite{PhysRevD.47.1497}.
If we expand the equation for $r \rightarrow +\infty$, we define $\epsilon_{S,B} = \omega r_{S,B}$ and we perform a change of variable $z = \omega r$, we get 
\begin{align}
    &\biggl[ 1 - \frac{\epsilon_B + 2 \epsilon_B}{z} +\frac{2 \epsilon_S \epsilon_B + \epsilon_S^2}{z^2} -\frac{\epsilon_B \epsilon_S^2}{z^3} + \mathcal{O}\biggl( \frac{1}{z^4} \biggr) \biggr]\frac{d^2 \Psi(z)}{d z^2} \nonumber \\
    &+ \biggl[ \frac{\epsilon_S + \epsilon_B/2}{z^2} - \frac{2\epsilon_S \epsilon_B + \epsilon_S^2}{z^3} + \mathcal{O}\biggl( \frac{1}{z^4} \biggr) \biggr] \frac{d \Psi(z)}{d z} \nonumber \\
    & \biggl[ 1 - \frac{l(l+1)}{z^2} +\frac{\epsilon_S(l^2+l-1) - \epsilon_B/2}{z^3} + \mathcal{O}\biggl( \frac{1}{z^4} \biggr) \biggr]\Psi(z) = 0 \,,
\end{align}
that in the limit $\{\epsilon_S,\epsilon_B\} \ll 1$ simply becomes
\begin{equation}\label{eq:Bessel_sol}
    \frac{d^2 \Psi(z)}{dz^2} + \left[ 1 - \frac{l(l+1)}{z^2} \right]\Psi(z)=0 \,.
\end{equation}
The general solution of the above equation is a linear combination of Riccati-Bessel functions $\sqrt{z} J_{l+\frac12}(z)$ and $\sqrt{z} N_{l+\frac12}(z)$. The solution $\Psi_-$ must be regular at the inner boundary, therefore we can choose just Bessel functions of the first kind, namely 
\begin{equation}\label{eq:Bessel_sol_small_z}
    \Psi_-(z) = B \sqrt{z} J_{l+\frac12}(z)\,,
\end{equation}
where $B$ is a constant. By taking the small $z$ expansion of the Bessel function we get
\begin{align}\label{eq:sol_inf_z_small}
    \Psi_-(z \ll 1) &= \frac{B}{2^{l+1/2}\Gamma(l+3/2)}z^{l+1} + \mathcal{O}(z^2) \nonumber \\
    & =\frac{B \sqrt{2} \, \epsilon_S^{l+1}}{(2 r_S)^{l+1} \Gamma(l+\frac32)}r^{l+1} + \mathcal{O}(r^2) \,.
\end{align}
In order to find the value of the constant $B$, we match the two solutions computed near the inner boundary and at infinity and stretched for large and small values of $r$, respectively. We get
\begin{equation}
    B = \frac{(2 r_S)^{l+1} \Gamma(l+3/2)\Gamma(1+2\alpha)\Gamma(1-2\beta)}{\sqrt{2} \epsilon_S^{l+1} r_S^{3/4} \, (r_S-r_B)^{l+1/4} \, \Gamma(1+\alpha-\beta)^2} \,.
\end{equation}
The amplitude $A_{\rm in}$ that appears in the Wronskian can be computed by comparing the far-field solution with the asymptotic behavior of Eq.~\eqref{eq:sol_inf_BC}
\begin{equation}
    A_{\rm in} = \frac{i^{l+1} (2r_S)^{l+1} \Gamma(l+3/2) \Gamma(1+2\alpha) \Gamma(1-2\beta)}{2 \sqrt{\pi} \epsilon_S^{l+1} r_S^{3/4} \, (r_S-r_B)^{l+1/4} \, \Gamma(1+\alpha - \beta)^2} \,.
\end{equation}
Using Eqs.~\eqref{eq:sol_inf_z_small},~\eqref{eq:scal_sol_inf} and~\eqref{eq:flux_inf} we get the $lm-$component of the energy flux at infinity: 
\begin{equation}\label{eq:lm_analytical_flux}
   \Dot{E}_\infty^{lm} =  \biggl( \frac{\mu \, q}{R_y} \biggr)^2 \biggl| \frac{\sqrt{\pi} \, Y_{lm}^*\left(\frac{\pi}{2},0\right)}{2^{l+1} \Gamma(l+\frac32)}  \biggr|^2 m^{2l+2} \left( \frac{r_S}{2} \right)^{l+1} r_0^{-(l+3)} \,,
\end{equation}
where we used $\omega = m \Omega$ and the fact that at large distances $f_B \rightarrow 1$ and $u^t \rightarrow 1$.  
Note that in the Schwarzschild limit the energy flux correctly reduces to the one of a scalar particle in circular orbit around the Schwarzschild BH~\cite{Brito:2012gw}, after identifying $q/R_y$ with the scalar charge.

A similar procedure can be followed to compute the analytical approximation of the scalar energy flux at $r_S$. 
In this case, the solution to Eq.~\eqref{eq:Bessel_sol} with the appropriate boundary conditions is given by 
\begin{equation}
    \Psi_+(z) = C \sqrt{z} \, H_{l+1/2}^{(1)}(z) \,,
\end{equation}
where $H_{l+1/2}^{(1)}(z) = J_{l+1/2}(z) + i N_{l+1/2}(z)$ is the Hankel function of the first kind. 
By matching this asymptotic solution with the correct boundary condition at infinity 
\begin{equation}
    \Psi_+(z \rightarrow +\infty) \sim e^{iz} \,,
\end{equation}
and using the expression of the Hankel function for large argument
\begin{equation}
    C \, \sqrt{z} \, H_{l+1/2}^{(1)}(z \rightarrow +\infty) \sim C \sqrt{\frac{2}{\pi}} e^{i z} (-i)^{l+1} \,,
\end{equation}
we get $C = i^{l+1} \sqrt{\frac{\pi}{2}}$.
If we consider the small-argument behavior of the Riccati-Bessel functions 
\begin{align}
    J_{l+1/2}(z \ll 1) & \sim \frac{z^{l+1/2}}{2^{l+1/2} \, \Gamma(l+3/2)} [1+ \mathcal{O}(z^2)] \,, \nonumber \\
    N_{l+1/2}(z \ll 1) & \sim - \frac{2^{l+1/2}}{\pi} \frac{\Gamma(l+1/2)}{z^{l+1/2}}[1+\mathcal{O}(z^2)] \,,
\end{align}
we note that the function $J_{l+1/2}$ is subdominant, therefore we get
\begin{equation}
    \Psi_+(z \ll 1) \sim i^l \frac{2^l}{\sqrt{\pi}} \Gamma(l+1/2)z^{-l} \,.
\end{equation}
Finally, using Eqs.~\eqref{eq:scal_sol_hor} and~\eqref{eq:flux_inf} we can compute the $lm$-component of the flux at the horizon
\begin{align}\label{eq:lm_analytical_flux_H}
    \Dot E_{r_S}^{lm} =  \biggl( \frac{\mu q}{R_y} \biggr)^2 &\biggl| Y_{lm}^*\biggl( \frac{\pi}{2},0 \biggr) \biggr|^2 \Bar{\Gamma}^2 \nonumber \\
    &\times \frac{m^2 r_S^{5/2} (r_S - r_B)^{2l+1/2}}{2} r_0^{-2l-5} \,,
\end{align}
where we defined 
\begin{equation}
    \Bar{\Gamma} \equiv \frac{\Gamma(l+1/2) \Gamma(1+\alpha - \beta)^2}{2 \, \Gamma(l+3/2) \Gamma(1+2\alpha) \Gamma(1-2\beta)} \,.
\end{equation}

\subsection{TS}
For the study of the analytical solution in the case of the TS we follow a procedure similar to the one adopted in the previous section, with some notable differences due to the fact that now the inner boundary is at radial distance $r_B$. For this reason we define the variable $h$ as follows
\begin{equation}
    h \equiv \frac{r-r_B}{r-r_S} \,.
\end{equation}
The radial equation for $h \ll 1$ gives
\begin{align}
    & h(1-h)^2 \frac{d^2 \psi}{dh^2} +(1-h) \frac{d\psi}{d h} \nonumber \\
    & + \biggl[ \frac{r_B^3 \omega^2}{(1-h)(r_B-r_S)}  - \frac{l(l+1)}{1-h} \biggr] \psi = 0 \,.
\end{align}
Using the field redefinition $\psi(h) \equiv h^\alpha (1-h)^\beta F(h)$ we get that the hypergeometric equation~\eqref{eq:hypergeom}, where again $a = b = \alpha + \beta$ and $c = 1+2 \alpha$, but this time $\alpha$ and $\beta$ are solutions to
\begin{align}
    \alpha^2 &= 0\,, \qquad
    \beta^2 - \beta + \frac{r_B^3 \omega^2}{r_B - r_S}  -l(l+1) &= 0\,,
\end{align}
namely,
\begin{align}
    \alpha &= 0\,,\qquad 
    \beta_{\pm} &= \frac12 \left[1 \pm \sqrt{(1+2l)^2 + 4\frac{r_B^3 \omega^2}{r_B - r_S}}\right] \,.
\end{align}
Since in this case $c=1$, the general solution of the hypergeometric equation is \cite{abramowitz+stegun}
\begin{align}\label{eq:hypergeo_TS}
    \psi (h) = &A_1 (1-h)^\beta F(a,b,1;h) + \nonumber \\
    &B_1 (1-h)^\beta \biggl\{\ln{|h|F(a,b,1;h)} 
    + \sum_{n=1}^\infty \frac{(a)_n (b)_c}{(n!)^2} h^n \nonumber \\
    &\times [\chi(a+n) -\chi(a) +\chi(b+n) - \chi(b) \nonumber \\
   &\hspace{2cm}-2\chi(n+1) + 2\chi(1)]
    \biggr\}\,,
\end{align}
where $\chi$ is the digamma function, i.e. the logarithmic derivative of the Gamma function. Since we consider the approximation $h \ll 1$ we can neglect the $\mathcal{O}(h)$ terms within the curly brackets. Furthermore, to ensure that the boundary condition of the TS near the inner boundary are satisfied, we require $B_1 = 0$. Using the properties of the hypergeometric functions, we get
\begin{align}
    \psi(h) &= A_1 (1-h)^\beta \biggl[ \frac{\Gamma(-2\beta)}{\Gamma(\beta)^2} F(a,b,a+b;1-h) \nonumber \\ 
    &+ (1-h)^{-2\beta} \frac{\Gamma(2\beta)}{\Gamma(\beta)^2} F(-a,-b,-a-b,1-h) \biggr]\,.
\end{align}
In the limit $1-h \ll 1$ and within the low-frequency approximation,  the above solution becomes 
\begin{align}
    \Psi(r) &= r f_B^{1/4} \psi (r) \sim \frac{r^{l+1}}{r_B^{3/4}(r_B - r_S)^l} \frac{\Gamma(-2\beta)}{\Gamma(1-\beta)^2} \,,
\end{align}
where we derived $A_1 = r_B^{3/4}$ from the comparison between Eq.~\eqref{eq:hypergeo_TS} with $B_1=0$ and the boundary condition of Eq.~\eqref{eq:BC_rb_TS}.
The far-field solution is identical to the one of the magnetized BH, see Eq.~\eqref{eq:Bessel_sol}, that for small $z$ becomes Eq.~\eqref{eq:Bessel_sol_small_z} with 
\begin{equation}
    B = \frac{(2 r_S)^{l+1} \Gamma(l+3/2) \Gamma(-2\beta)}{\sqrt{2} \epsilon_S^{l+1} r_B^{3/4} \, (r_B - r_S)^l \Gamma(1-\beta)^2} \,.
\end{equation}
The amplitude $A_{\rm in}$ is instead given by 
\begin{equation}
    A_{\rm in} = \frac{i^{l+1} (2r_S)^{l+1} \Gamma(l+3/2) \Gamma(-2 \beta)}{2 \sqrt{\pi} \epsilon_S^{l+1} r_B^{3/4} \, (r_B - r_S)^l \Gamma(1-\beta)^2} \,.
\end{equation}
It turns out that the $lm-$component of the energy flux at infinity is given again by Eq.~\eqref{eq:lm_analytical_flux}.
The fact that the analytical expression for the flux at infinity is the same for both BH and TS is a consequence of the fact that the two solutions coincide asymptotically and their different inner boundaries are negligible in the small-frequency regime. Since the low frequency approximation holds for large orbital radii, the analytical flux at infinity is expected to be the same.

\section{Numerical results}
This section presents the numerical results for the scalar energy flux and its comparison with the low-frequency analytical approximation in the case of both magnetized BHs and TSs. 
The emitted power is computed for different values of the ratio $r_B/r_S$ and varying the radial distance of the stable circular orbit $r_0>r_{\rm ISCO}$.

For the magnetized BH, the general behavior of the flux at infinity is overall quite similar to the one of a Schwarzschild BH. 
Although this is true also for first kind TSs, the case of second kind TSs is particularly interesting since, for ultracompact solutions (i.e. when $r_B/r_S\to1$), the orbiting scalar particle can excite long-lived QNMs with a low real frequency, leading to sharp resonances in the emitted energy flux that are not present in standard BHs within GR.

Finally, we also compute the dephasing between the magnetized BH and the TS, accumulated within one year of evolution of the test scalar charge up to the ISCO of the BH. 

\subsection{Energy flux of magnetized BHs and TSs}

The scalar energy fluxes are computed using Eq.~\eqref{eq:flux_inf} and numerically solving the source free version of Eq.~\eqref{eq:Schro_eq} with the correct boundary conditions to get $\Psi_-$. The fluxes are compared to the analytical result which is valid at large orbital separation (see Sec.~\ref{sec:analytical_solution}). 
As already mentioned, in the low-frequency approximation, the $lm-$ component of the flux at infinity is described by Eq.~\eqref{eq:lm_analytical_flux} for both magnetized BHs and TSs, while the flux at the horizon is given by Eq.~\eqref{eq:lm_analytical_flux_H} for BHs.

\begin{figure}[th]
  \centering
\includegraphics[width=0.485\textwidth]{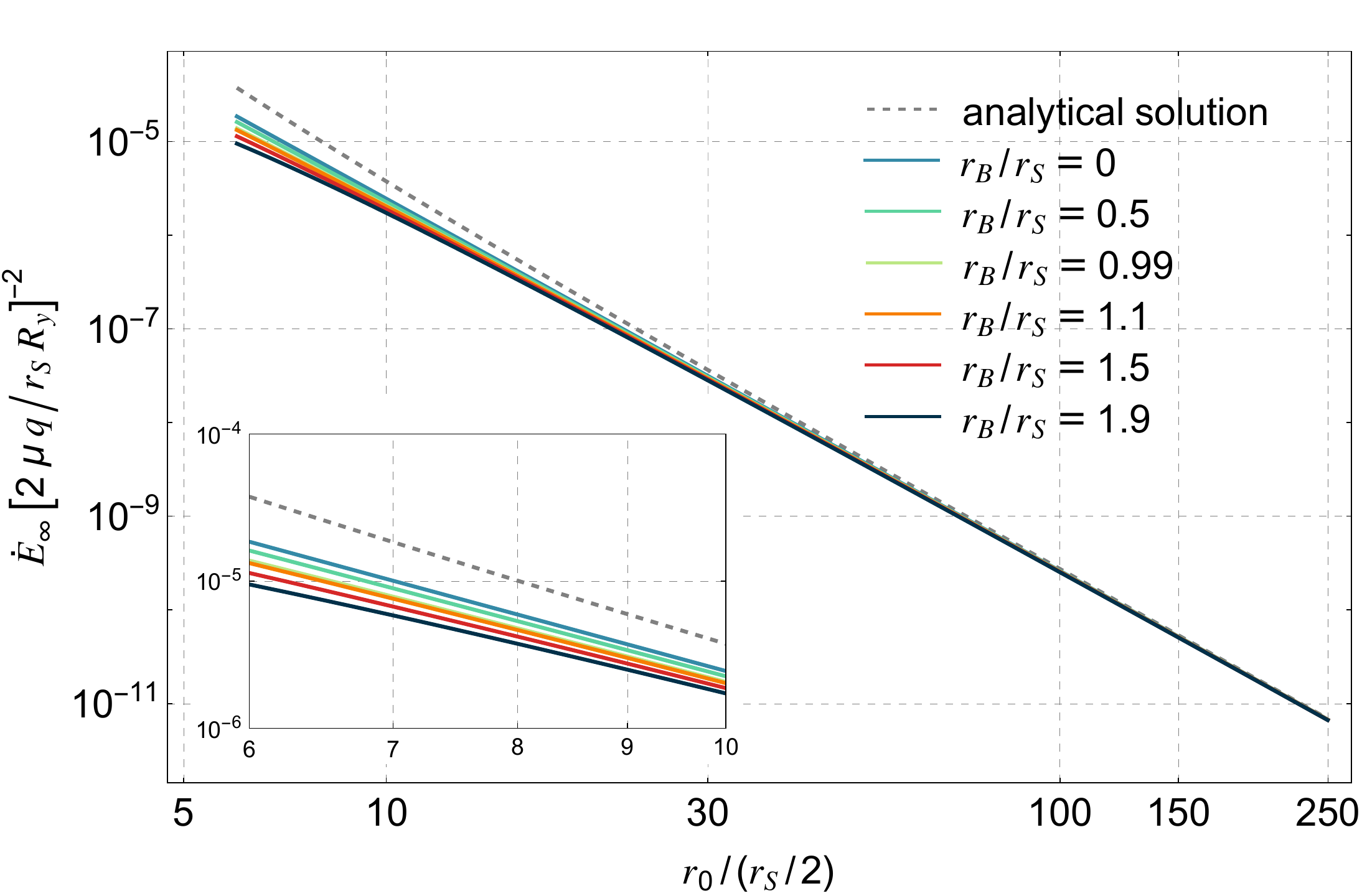}
  \caption{Scalar energy flux at infinity as a function of the orbital radial distance $r_0$, starting from $r_{\rm ISCO}$, for different values of the ratio $r_B / r_S$ of the magnetized BH and the TS. 
  The flux is computed truncating the summation at $l_{\rm max} = 3$.
  The dashed gray line represents the flux computed using the analytical solution.
  The inset shows a zoom of the region close to the ISCO, where the differences in the flux are larger.
  }\label{fig:flux_inf_MBHandTS_rS}
\end{figure}

\begin{figure}[th]
  \centering
\includegraphics[width=0.485\textwidth]{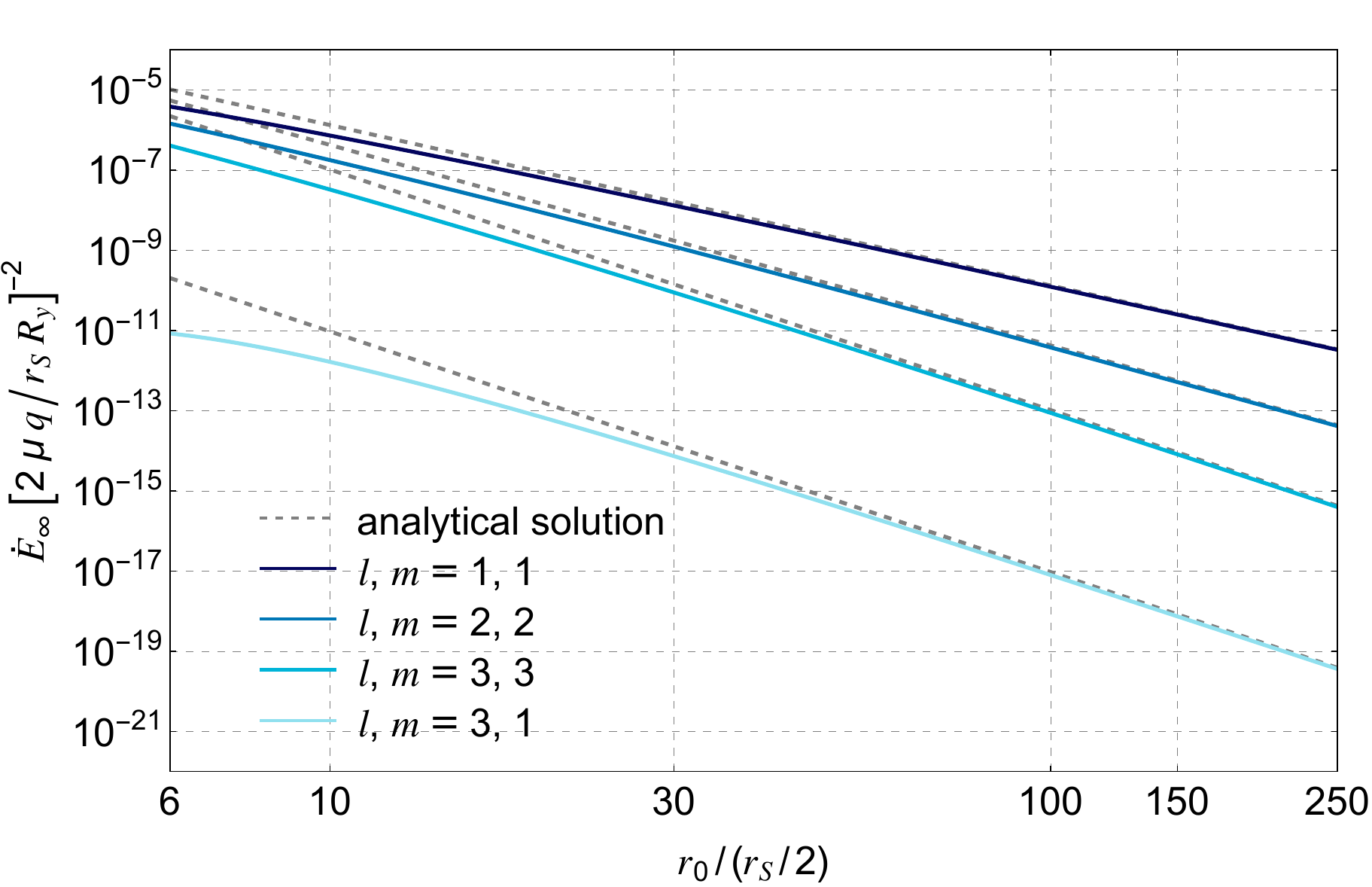}
  \caption{$lm$-component of the scalar energy flux emitted at infinity as a function of the radial distance $r_0$ of the circular orbit, starting from $r_{\rm ISCO}$, for a TS with $r_B / r_S = 1.5$.
  The dashed gray lines represent the flux computed using the low-frequency analytical approximation. Note that the cases $l,m = 2,1$ and $l,m = 3,2$ are not present because the spherical harmonics $Y_{21}(\pi/2,0)=Y_{32}(\pi/2,0)=0.$
  }\label{fig:loglog_lm_flux_inf_TS_rS}
\end{figure}

The results are shown in Fig.~\ref{fig:flux_inf_MBHandTS_rS} for both magnetized BHs and TSs and different values of the parameters ratio $r_B/r_S$.
The top panel shows good agreement between the numerical results for $r_0\gg r_S$ and the analytical approximation, which gradually loses validity as the orbit approaches the ISCO, as evident from the bottom panel.
This agreement is present also for each $lm$-component of the flux, as shown in Fig.~\ref{fig:loglog_lm_flux_inf_TS_rS} for a TS.
As expected and partially confirmed by the analytical expression, the maximum power emitted at infinity occurs when the particle orbits along the ISCO and monotonically decreases for circular orbits farther from the compact object.

The total flux is computed by truncating the $lm$ summation to $l_{\rm max}=3$, since the other contributions are subleading and their inclusion does not significantly affect the results we wish to discuss.
Indeed, as can be seen from Fig.~\ref{fig:loglog_lm_flux_inf_TS_rS} for a TS with $r_B/r_S = 1.5$, the leading term is given by the dipolar ($l=1$) mode and each subsequent contribution decreases monotonically as $l$ increases, also in agreement with the analytical expression in Eq.~\eqref{eq:lm_analytical_flux}.

\begin{figure}[th]
  \centering
\includegraphics[width=0.485\textwidth]{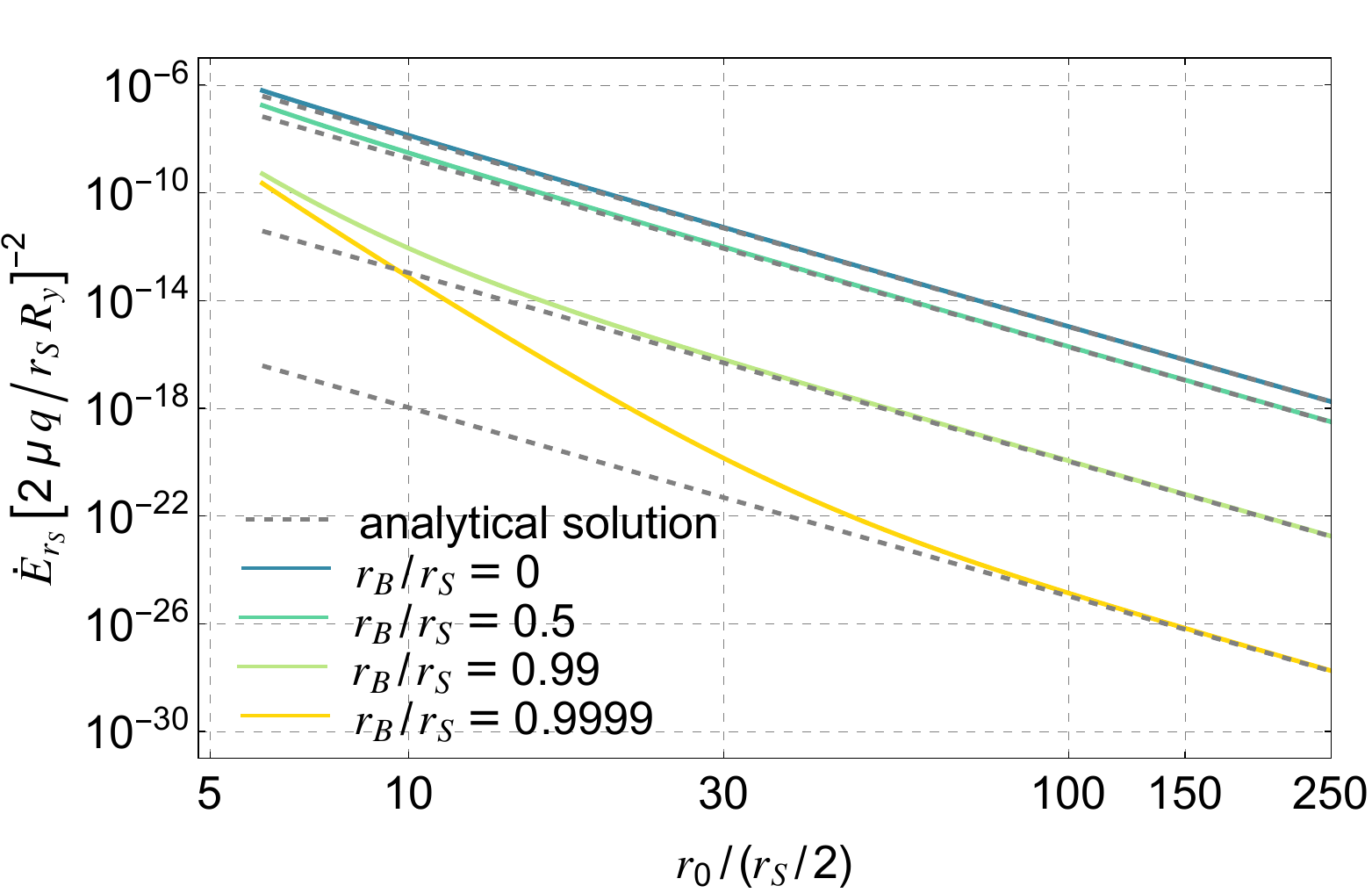}
  \caption{Scalar energy flux at the horizon as a function of the orbital distance $r_0$, starting from $r_{\rm ISCO}$, for different values of the ratio $r_B / r_S$ of the magnetized BH. 
  The flux is computed truncating the summation at $l_{\rm max} = 3$.
}\label{fig:flux_h_MBH_rS}
\end{figure}

Due to the presence of the horizon at $r_S$ in the magnetized BH case, we also computed the scalar flux absorbed by the BH, as shown in Fig.~\ref{fig:flux_h_MBH_rS}. 
The comparison between the numerical results and the analytical solution of Eq.~\eqref{eq:lm_analytical_flux_H} provides an excellent agreement for large orbital separation.
In all cases, the contribution of the flux at the horizon is subleading compared to the flux at infinity.
As can be seen from the plot, its value decreases as the BH approaches extremality at $r_B/r_S = 1$, as expected, since this is the point in parameter space where the BH and the TS are continuously connected.
As opposed to the flux at infinity, the fluxes at the horizon for the different $r_B/r_S$ values considered do not asymptotically tend to the same value for large orbital radii.

\subsection{Resonances}
A particularly intriguing aspect in the analysis of the energy flux emitted at infinity concerns ultra-compact TSs of the \textit{second kind}, which behave like BH mimickers but exhibit key differences from BHs that can serve as signatures of the absence of the horizon~\cite{Cardoso:2019rvt}. 
As can be seen from the expression of the source in Eq.~\eqref{eq:source_eq}, the point particle following a stable circular orbit around the primary object can excite the QNMs of the compact object as long as $m \Omega$ matches the real part of the QNMs frequency
\begin{equation}\label{eq:resonant_condition}
    \omega_R = m\Omega \,,
\end{equation}
where we defined $\omega_R={\rm Re}[\omega_\text{QNM}]$ and $\omega_I={\rm Im}[\omega_\text{QNM}]$.
When the resonant condition above is fulfilled, the energy flux develops sharp peaks for values of $r_0$ that satisfy Eq.~\eqref{eq:resonant_condition}. 
This is clear also from the denominator of Eq.~\eqref{eq:scal_sol_inf} that involves the Wronskian, which vanishes when the considered frequency corresponds to the one of a QNM. If the modes are long-lived ($|\omega_I|\ll\omega_R$, as it happens for ultracompact TSs~\cite{Heidmann:2023ojf,Dima:2024cok,Bena:2024hoh,Dima:2025zot}) then the Wronskian is almost zero on the real axis when the resonant condition~\eqref{eq:resonant_condition} is satisfied.
For BHs in 4 dimensional vacuum GR, this never occurs, as the real part of the fundamental QNM frequency is always higher than the orbital frequency of the ISCO for both static and rotating BH solutions. 
On the other hand, the QNM spectrum of ultracompact horizonless objects contains low-frequency modes trapped within the object and the photon-sphere barrier~\cite{Cardoso:2014sna,Cardoso:2019rvt}, and these modes can be resonantly excited by circular orbits.

\begin{figure}[th]
  \centering
\includegraphics[width=0.485\textwidth]{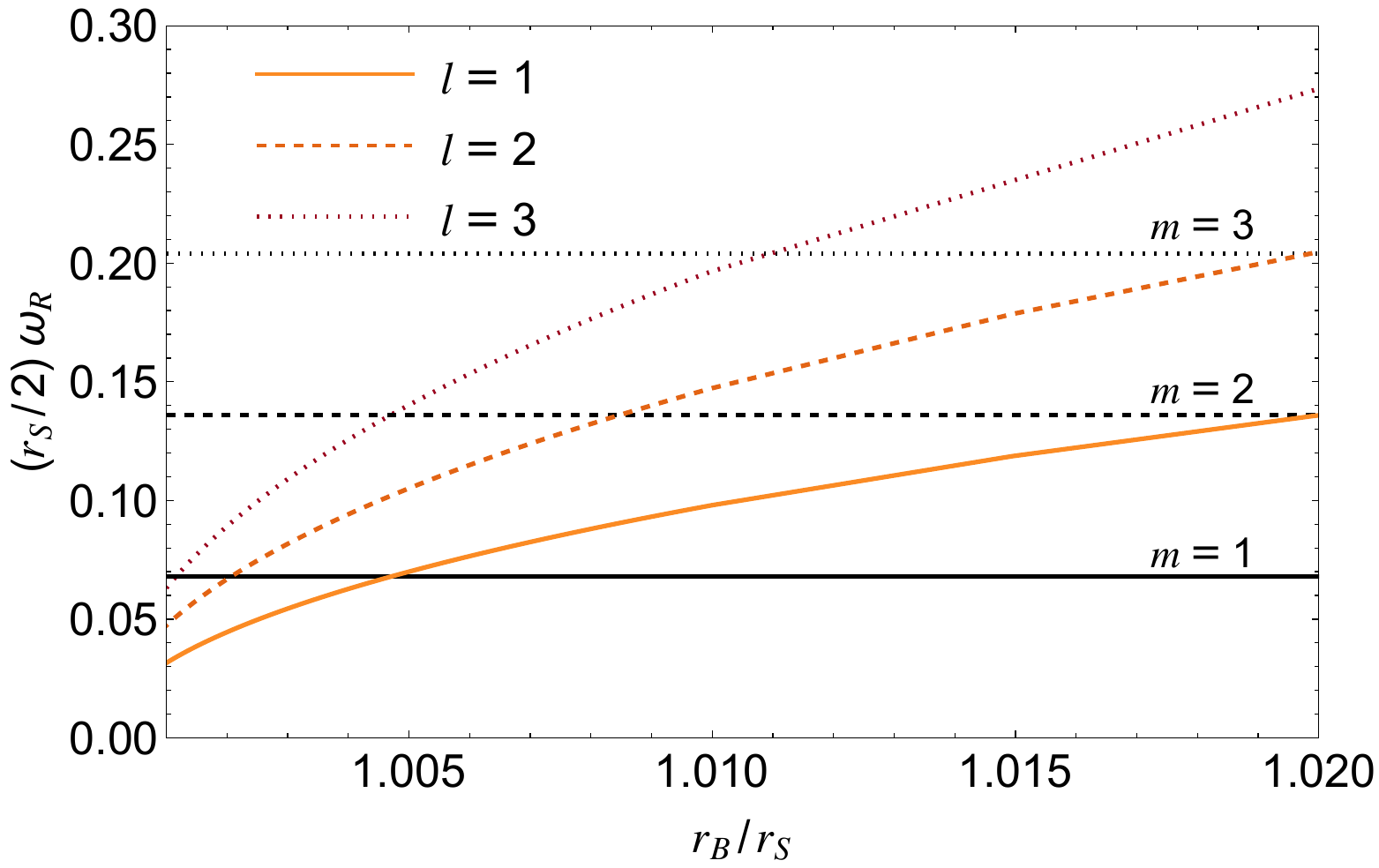}
  \caption{Real part of the QNM frequencies for test scalar field perturbations in the background of a TS, for different values of $l = 1, 2, 3$. 
  The solid, dashed and dotted horizontal lines represent multiples of the orbital frequency at the ISCO, $m \, \Omega_{\rm ISCO} \, (r_S / 2)$ for $m=1,2,3$, respectively.
  }\label{fig:real_QNMs}
\end{figure}

\begin{figure}[th]
  \centering
\includegraphics[width=0.485\textwidth]{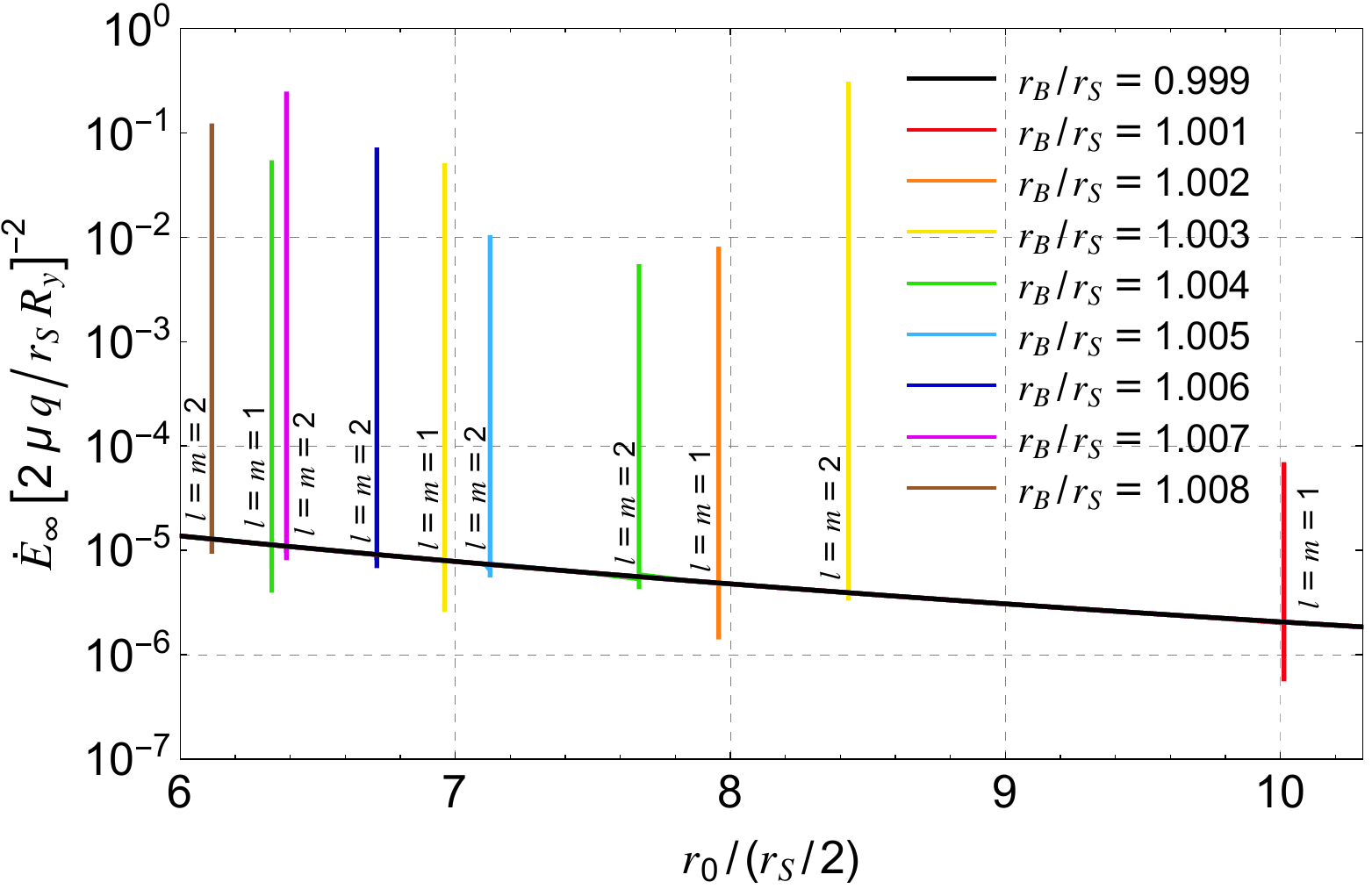}
  \caption{ 
    Resonances of the scalar energy flux at infinity for different values of $r_B/r_S$ in the parameter space of ultracompact TSs. The black curve represents the scalar energy flux at infinity for a nearly-extremal magnetized BH, providing a comparison with the considered TS solutions. For each resonance, we indicate the angular momentum and azimuthal numbers of the excited QNM that generates the peak in the flux.  }\label{fig:all_resonances_rS}
\end{figure}

The QNMs of a test scalar field in the background of a TS in the full parameter space are shown in Appendix~\ref{app:QNMs}. 
The TS presents long-lived modes with sufficiently small real frequencies for small enough values of $r_B/r_S$, as shown in Fig.~\ref{fig:real_QNMs}. 
Indeed, as long as the QNM real frequency remains below the horizontal lines (which represent multiples of the ISCO orbital frequency, $m\Omega_{\rm ISCO}$) the point particle can excite the corresponding QNM. 
By exploring the relevant parameter space, guided by the information of Fig.~\ref{fig:real_QNMs}, we found several resonances in the energy flux of the TS, for different values of $r_B/r_S$. 
The results are summarized in Fig.~\ref{fig:all_resonances_rS}, where the scalar energy flux at infinity for several TSs with different $r_B/r_S$, along with their characteristic resonances, is compared to that of a nearly-extremal magnetized BH. 
The latter does not exhibit any peaks, since its modes are neither long-lived nor have sufficiently low frequency.
The resonances presented in the plot occur for excited QNMs with $l=m=1$ and $l=m=2$, where the value of $m$ enters the resonant condition Eq.~\eqref{eq:resonant_condition} and the flux formula in Eq.~\eqref{eq:flux_inf}, while the QNMs are degenerate in the azimuthal number, due to the spherical symmetry of the background.

\begin{figure}[th]
  \centering
\includegraphics[width=0.485\textwidth]{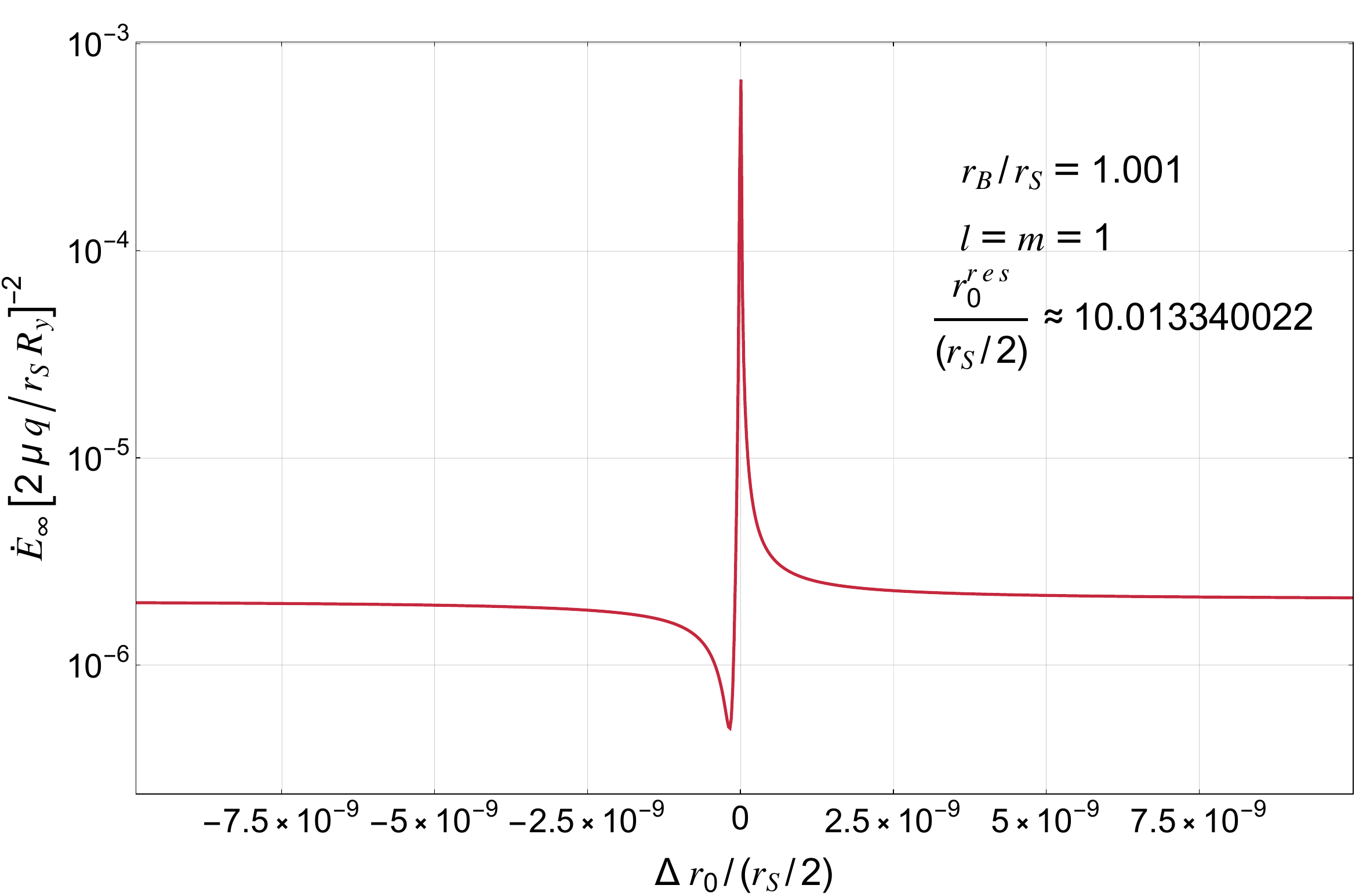}
\includegraphics[width=0.485\textwidth]{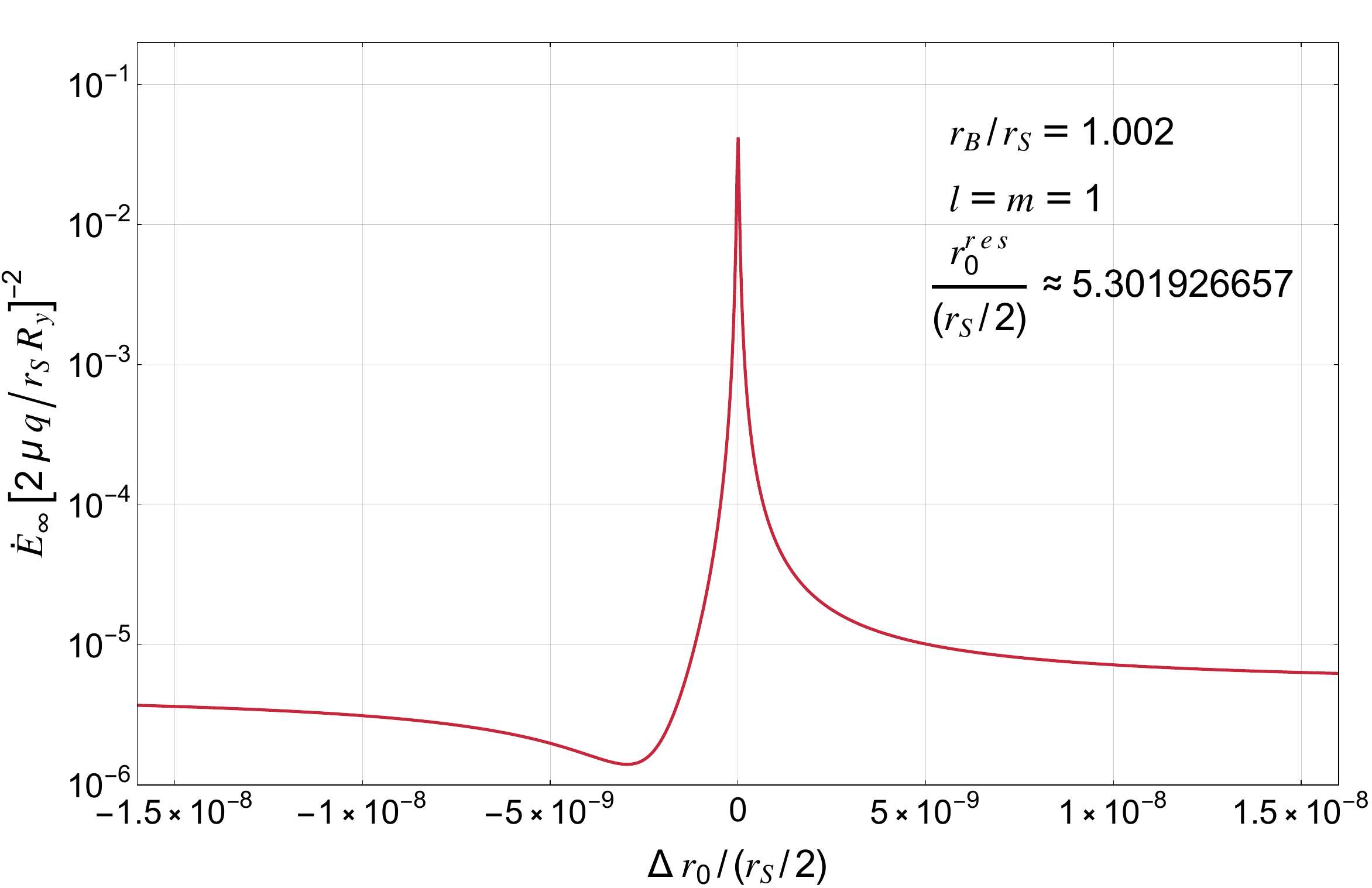}
\includegraphics[width=0.485\textwidth]{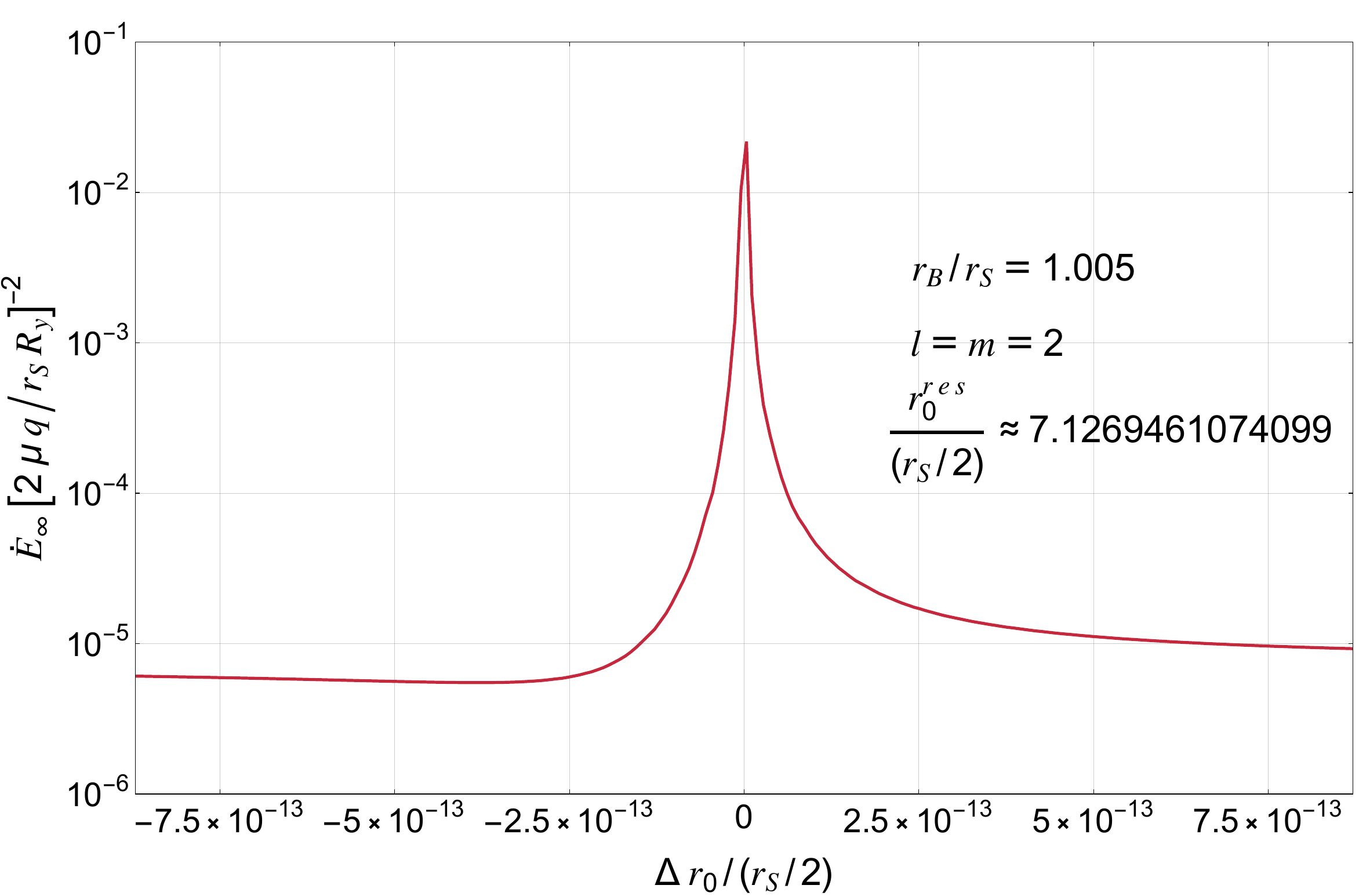}
  \caption{Resonances of the scalar energy flux at infinity for TSs of the \textit{second kind} with $r_B/r_S = 1.001$ (top panel), $r_B/r_S = 1.002$ (middle panel) and $r_B/r_S = 1.005$ (bottom panel). 
  The peaks are associated to QNMs with $l=m=1$ (top and middle panels) and $l=m=2$ (bottom panel). 
  The origin of the axis is chosen to coincide with the radial distance $r_0^{\rm res}$, where the peak of the resonance occurs.
  }\label{fig:specific_resonances_rS}
\end{figure}

A clearer view at some specific resonances is present in Fig.~\ref{fig:specific_resonances_rS} where, zooming-in near a resonance, we show the typical structure of the peaks for three distinct TSs. 
Note that, in all cases, the resonances are extremely narrow, and we needed to employ high resolution in scanning the orbital distance to identify them. 
This is due to the rather small value of the imaginary part of the excited QNM frequency, that can range from $(r_S/2)|\omega_I| \sim 10^{-11}$ for the $l=1$ mode of the TS with $r_B/r_S = 1.004$, to $(r_S/2)|\omega_I| \sim 10^{-19}$ for the $l=2$ mode of the TS with $r_B/r_S = 1.003$.
The other QNMs with $(l,m)=(2,2), (3,3), (3,1)$ that, according to Fig.~\ref{fig:real_QNMs}, are prone to be excited were not identified because of their extremely small imaginary part ($(r_S/2) |\omega_I| \lesssim 10^{-20}$, as shown in Appendix~\ref{app:QNMs}).

\begin{figure}[th]
  \centering
\includegraphics[width=0.485\textwidth]{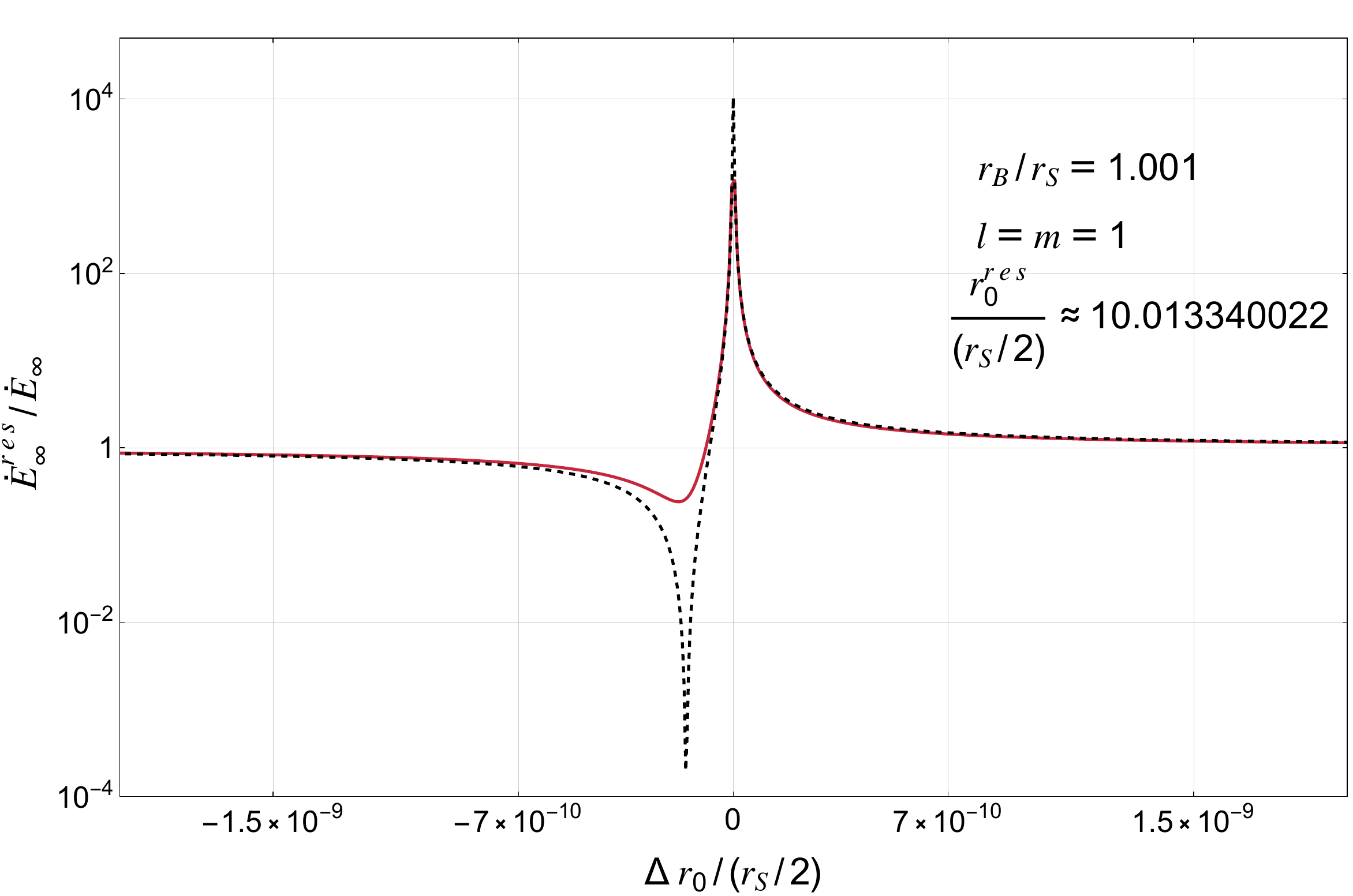}
  \caption{
  Comparison between the energy flux at infinity $\Dot{E}_\infty^{\rm res}$, normalized by the flux ignoring the resonance $\Dot{E}_\infty$, computed with the exact numerical results (red solid line) and with the best fit of a forced-oscillator model described by Eq.~\eqref{eq:fit_resonance}  (dashed black line).
  }\label{fig:fit_res_comparison_rS}
\end{figure}

The minimum shown in the plots, often referred to as an anti-resonance, is common and can be easily explained by a simple toy model where the perturbations of the compact object are described by a forced harmonic oscillator, sourced by the secondary object (see \cite{Pons:2001xs,Maggio:2021uge,Pani:2010em} for more details).
This allows us to model the energy flux across a single resonance as
\begin{equation}\label{eq:fit_resonance}
    \frac{\Dot{E}_\infty^{\rm res}}{\Dot{E}_\infty} = \frac{[(1-b)(m\Omega)^2 - \omega_R^2 - \omega_I^2]^2+(2m\Omega \omega_I)^2}{{[(m\Omega)^2 - \omega_R^2 - \omega_I^2]}^2+(2m\Omega \omega_I)^2} \,,
\end{equation}
where $\Dot{E}_\infty^{\rm res}$ and $\Dot{E}_\infty$ are the energy flux at infinity with and without the resonance, respectively, $b = 1 - (\Omega_{\rm max}/\Omega_{\rm min})^2$, with $\Omega_{\rm max}$ and $\Omega_{\rm min}$ denoting the frequencies corresponding to the maximum and the minimum of the resonance. Figure~\ref{fig:fit_res_comparison_rS} shows the resonance for a TS with $r_B/r_S = 1.001$, associated to the $l=m=1$ mode, together with the best fit obtained using Eq.~\eqref{eq:fit_resonance}. 

\subsection{Dephasing}

To close this section we now compute the phase of the scalar wave emitted by the test charge around either a TS or a magnetized BH, focusing on the difference between these two situations in terms of the dephasing. 
We therefore define
\begin{equation}\label{eq:dephasing_eq}
    \delta \phi_s (t) = | \phi_s^{\rm BH}(t) - \phi_s^{\rm TS}(t)|\,,
\end{equation}
where $\phi_s^{\rm BH}(t)$ and $\phi_s^{\rm TS}(t)$ are the phase at time $t$ for the BH and TS, respectively. Without loss of generality we can choose $\phi_s^{\rm BH}(t=0) = \phi_s^{\rm TS}(t=0) = 0$.

To estimate the contribution of the resonances to the evolution of the inspiral we computed the dephasing between the TS with the resonance and TS in which the resonance was artificially removed (by setting a resolution larger than the width of the resonance), for the case $r_B/r_S = 1.001$ shown in Fig.~\ref{fig:fit_res_comparison_rS}. 
For this purpose it is useful to introduce the dephasing accumulated during the inspiral evolution from an initial radius $r_{\rm ini}$ at $t_{\rm ini} = 0$ to a final radius $r_{\rm fin}$, between the two TS cases with and without the resonance
\begin{equation}
    \Delta \phi_s = |\phi_{s, \, {\rm res}}^{TS}(r_{\rm fin}) - \phi_{s, \, {\rm no \,res}}^{TS}(r_{\rm fin})| \,,
\end{equation}
where again $\phi_{s, \, {\rm res}}^{TS}(r_{\rm ini})=\phi_{s, \, {\rm no \,res}}^{TS}(r_{\rm ini}) = 0$.
Considering the evolution centered on the resonance with a width of $(r_{\rm fin}-r_{\rm ini}) / (r_S/2) \approx 3 \times 10^{-8}$, we computed the dephasing $\Delta \phi_s  \lesssim \mathcal{O}(10^{-1})$ rads. Since the $l=m=1$ resonance for the case of the TS with $r_B/r_S = 1.001$ is the widest among all those studied and shown in Fig.~\ref{fig:all_resonances_rS}, we conclude that all the resonances are too narrow to produce a detectable dephasing.
Indeed, as a rough rule of thumb, a dephasing greater than $1\,{\rm rad}$ would substantially impact a matched-filter search, leading to a significant loss of detected events~\cite{Lindblom:2008cm}. We emphasize that this rule of thumb should be confirmed through more detailed, model-dependent analysis; for instance, correlations might still enable detection using incorrect models, though this would introduce systematic errors in the estimated parameters.

\begin{figure}[th]
  \centering
\includegraphics[width=0.485\textwidth]{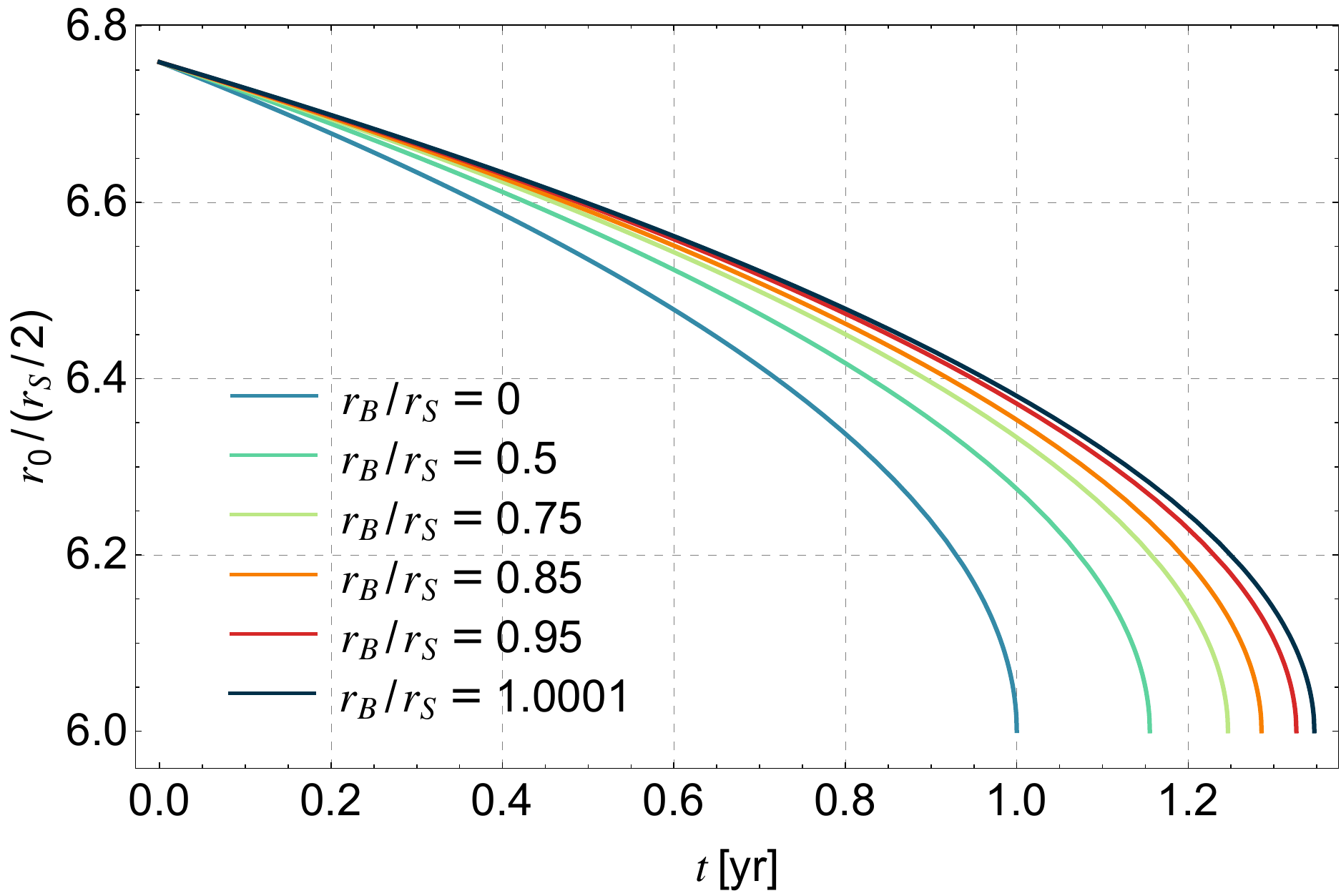}
  \caption{
  Evolution of the orbital radius of the point particle under radiation reaction for $q/R_y = 1$, $r_S/2 = 10^6 M_\odot$ and $\mu = 30 M_\odot$. The evolution begins at a radial distance $r_0 \approx 6.759 \, r_S/2$ at $t_{\rm ini} = 0$, to guarantee $1$ year of evolution for the magnetized BH case with $r_B/r_S = 0$, and stops as the particle reaches $r_{\rm ISCO}$ at $t_{\rm ISCO}$. 
  }\label{fig:time_ISCO}
\end{figure}

\begin{figure}[th]
  \centering
\includegraphics[width=0.485\textwidth]{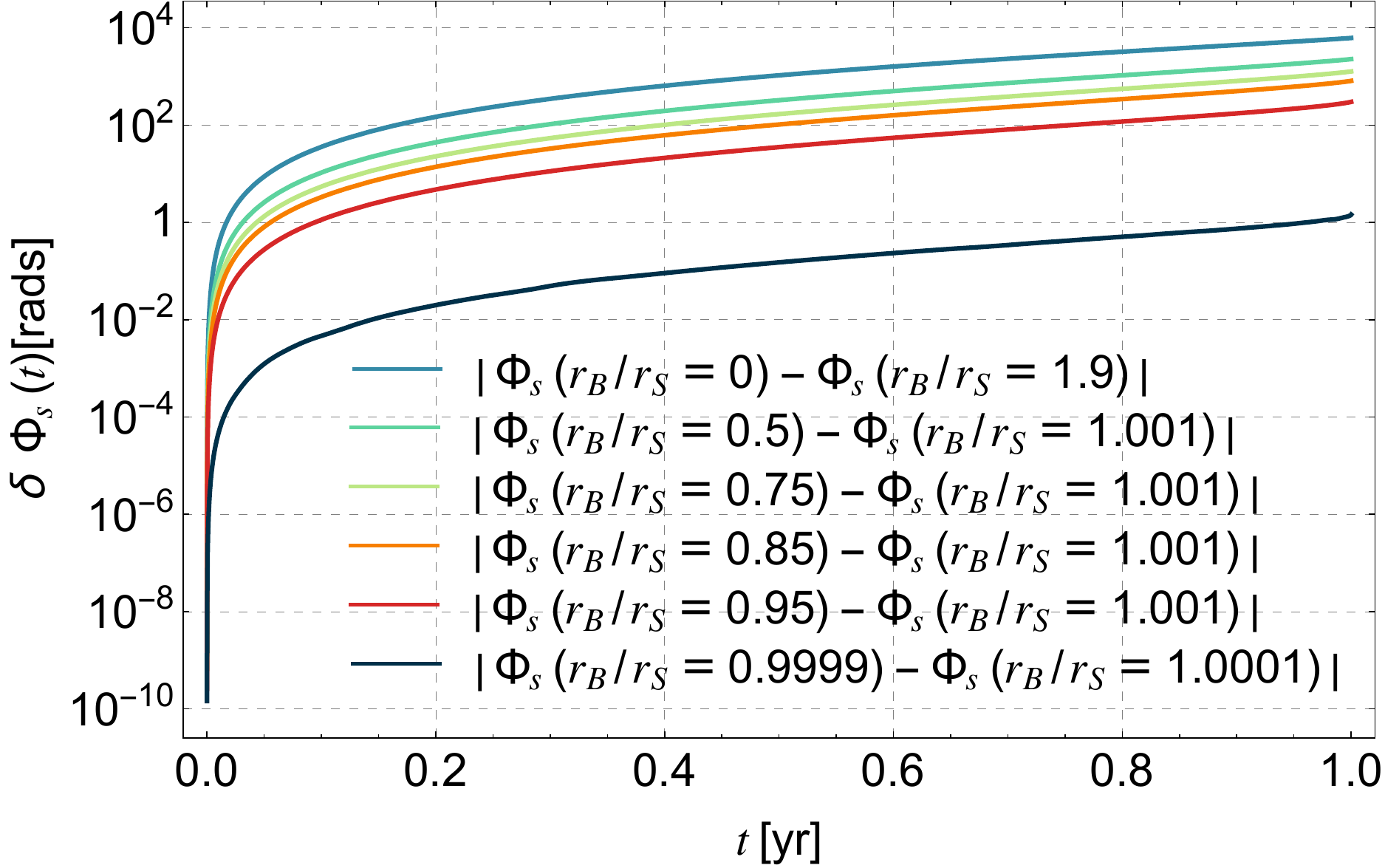}
  \caption{
  Scalar wave dephasing between magnetized BH and the TS as a function of time for $q/R_y = 1$, $r_S/2 = 10^6 M_\odot$ and $\mu = 30 M_\odot$. For each curve the starting point at $t_{\rm ini}=0$ is set to guarantee $1$ year of evolution up to the ISCO for the scalar charge orbiting around the  BH.
  }\label{fig:dephasing}
\end{figure}

For this reason we decided to neglect all the resonances and compute the dephasing between different magnetized BHs and TSs using Eq.~\eqref{eq:dephasing_eq}, as shown in Fig.~\ref{fig:dephasing} for parameters $q/R_y = 1$ \footnote{As discussed in the conclusion, the choice of $q/R_y=1$ is made to imitate the gravitational case, wherein the source is only proportional to $\mu$ and independent of $R_y$. Clearly the fluxes can be easily rescaled for different values of $q/R_y$.}, $r_S/2 = 10^6 M_\odot$ and $\mu = 30 M_\odot$.
In each case the starting point is chosen such that the test scalar charge completes the orbital evolution up to the ISCO of the BH in one year. 
Figure~\ref{fig:time_ISCO} shows the evolution of the orbital radius $r_0$ of circular orbits as a function of time, for both magnetized BHs and TSs, varying $r_B/r_S$. 
The test charge reaches the ISCO in the BH case faster than in the TS case (see Fig.~\ref{fig:time_ISCO}) due to the additional flux at the horizon and the larger flux at infinity, as shown in Fig.~\ref{fig:flux_inf_MBHandTS_rS}.
As Fig.~\ref{fig:dephasing} confirms, the further away the BH and the TS are in the parameter space, the larger is the accumulated dephasing, arriving up to $\mathcal{O}(10^4)$ rads in the comparison between the Schwarzschild limit and a TS with $r_B/r_S = 1.9$. 
On the other hand, we estimate that the minimum $\epsilon$ difference, defined as $r_B/r_S = 1 \pm \epsilon$, to develop a dephasing of around 1 radiant between a nearly extremal magnetized BH and an ultra-compact TS is $\epsilon \approx10^{-4}$, that corresponds to $\Delta Q/M \approx 4 \times 10^{-5}$ between the two solutions. 

As previously mentioned, the contribution of the scalar flux absorbed by the horizon in the BH case is always subdominant relative to the flux at infinity.
However, it is only near the extremal BH limit at $r_B/r_S = 1$ that it becomes completely negligible, as shown in Fig.~\ref{fig:dephasing}, which illustrates the dephasing after $1$ year of evolution between a BH with $r_B/r_S = 0.9999$ and a TS with $r_B/r_S = 1.0001$. 
The much larger dephasing observed in all other comparisons is due to the different fluxes, both at the horizon and at infinity.

\section{Conclusions \& Outlook}
We studied the radiation emitted by a test charge in circular orbit around magnetized TSs and BHs. The former are regular solutions to Einstein-Maxwell theory in five dimensions which are similar to BHs at large distance but replace the horizon by a regular interior.

The most interesting case is when the TS is compact enough to mimic a BH ($r_B\approx r_S$). When this happens, the low-frequency, long-lived QNMs of the TS can be excited during the inspiral. However, such resonances are too narrow to produce a significant effect. The main difference between the BH and the TS in the $r_B\to r_S$ limit is therefore due to the absence of energy flux at the horizon for a TS, as it happens in other toy models~\cite{Cardoso:2019nis,Datta:2019epe,Maggio:2021uge,Datta:2024vll}. For a TS, absence of effective trapping requires the condition~\eqref{trappingcond}. This condition is polynomial in $(r_B/r_S-1)$, while in effective toy models is typically logarithmic, $\Delta T\sim \log(r_B/r_S-1)$~\cite{Cardoso:2019rvt}, making effective trapping much less efficient than for a TSs~\cite{Datta:2019epe,Maggio:2021uge,Datta:2024vll}. In the TS case, assuming a mass ratio $\mu/M \approx 10^{-6}$, effective trapping would be efficient whenever
\begin{equation}
    r_B-r_S\lesssim \left(\frac{r_S}{10^6 M_\odot}\right)\,{\rm micron}\,.
\end{equation}
This difference is clearly tiny compared to the size of the primary, but still much larger than the Planck length, $\ell_{\rm Planck}$, that is typically considered as the relevant near-horizon length scale for quantum BH effects~\cite{Mathur:2009hf,Bena:2022rna,Bena:2022ldq}. In other words, if $r_B=r_S+\ell_{\rm Planck}$, the TS would be practically indistinguishable from a BH, at least in this model for what concerns radiation-reaction effects and QNM excitation.
It is also worth mentioning that, from a practical point of view, the dephasing between a TS with $r_B=r_S(1+\epsilon)$ and a BH with $r_S=r_B(1-\epsilon)$ is negligible when $\epsilon\lesssim 10^{-4}$, so the regime in which effective trapping can occur is likely beyond the observational accuracy.

We focused on scalar radiation by a test charge. The extension to the more relevant case of a point mass can be done by including a source into the analysis of~\cite{Dima:2024cok,Dima:2025zot}. This case is technically challenging because gravitational, scalar, and electromagnetic perturbations are all coupled to each other, but should be feasible with standard techniques.
In preparation of such a calculation, it is useful to estimate the scaling of the source with the model parameters. In the case of a point mass, the relevant action reads
\begin{align}
    S = \int d^5x \sqrt{-g} &\left( \frac{R}{2 \kappa_5^2} -\frac14 F_{\mu\nu}F^{\mu\nu}\right) - \mu  \int_\gamma d\tau \sqrt{g_{\mu\nu}u^\mu u^\nu} \,,
\end{align}
where $\gamma$ is the particle worldline.
This gives Einstein-Maxwell equations in five dimensions,
\begin{equation}
    G_{\mu\nu}=\kappa_5^2( T_{\mu\nu}^{\rm EM}+T_{\mu\nu}^{\rm particle})\,,
\end{equation}
where $T_{\mu\nu}^{\rm EM}$ is the electromagnetic stress-energy tensor and, for a particle in equatorial motion as described above,
\begin{equation}
    T_{\mu\nu}^{\rm particle} =\mu\frac{u_\mu u_\nu}{\sqrt{-g}u^t}\delta(r-r_0) \delta(\theta-\pi/2)\delta(\phi-\Omega t)\delta(y)\,.
\end{equation}
Following the same procedure done for the test scalar charge, one would obtain
\begin{align}
    T_{\mu\nu}^{\rm particle} =&\mu \frac{ u_\mu u_\nu}{r^2 u^t}\delta(r-r_0) 
    \sum_n \frac{e^{i n \frac{y}{R_y}}}{2 \pi R_y} \nonumber \\ 
    &\sum_{lm} Y_{lm}(\theta,\varphi) Y_{lm}^*\left(\frac{\pi}{2},0\right) e^{-im\Omega t}
    \,.
\end{align}
Interestingly, since $\kappa_5^2=\kappa_4^2 R_y$ and $T_{\mu\nu}^{\rm particle}\sim 1/R_y$, it is easy to see that $R_y$ disappears from the source of the massless ($n=0$) modes, which depends on $\kappa_5^2 T_{\mu\nu}^{\rm particle}$. This corresponds to the case $R_y=q$ previously discussed. We leave the full computation of the system's linear response to future work.

A more complex extension regards considering more realistic BH microstates~\cite{Bah:2022yji}, including non-extremal~\cite{Chakraborty:2025ger} or
spinning backgrounds. Solutions describing spinning TSs are recently under investigation~\cite{Bianchi:2025uis,private}. In case geodesic motion is integrable and perturbation equations can be separated for these solutions, as it happens (at least for test scalar perturbations) for the solution recently found in~\cite{Bianchi:2025uis}, this would enormously simplify the problem of computing the EMRI dynamics, as in the Kerr case. Otherwise, non-integrability of geodesic motion would result in chaotic behavior as for boson stars~\cite{Destounis:2023khj}, and absence of separability would require solving a system of partial differential equations to compute the emitted fluxes.
Nonetheless, both extensions (gravitational perturbations induced by a test mass and spinning backgrounds) are necessary to develop realistic waveform models for LISA data analysis~\cite{LISAConsortiumWaveformWorkingGroup:2023arg}.

\begin{acknowledgments} 
We are grateful to Iosif Bena, Massimo Bianchi, Donato Bini, Giorgio Di Russo, Pierre Heidmann, and Jorge Santos for fruitful interactions.
M.M. thanks CENTRA/IST for the kind hospitality during the realization of the project.
R.B. acknowledges financial support provided by FCT – Fundação para a Ciência e a Tecnologia, I.P., through the ERC-Portugal program Project ``GravNewFields''.
This work is partially supported by the MUR PRIN Grant 2020KR4KN2 ``String Theory as a bridge between Gauge Theories and Quantum Gravity'', by the FARE programme (GW-NEXT, CUP:~B84I20000100001), and by the INFN TEONGRAV initiative.
\end{acknowledgments}

\newpage

\appendix

\section{QNMs of a test scalar field in the background of the TS}
\label{app:QNMs}
In this Appendix, as a practical reference, we show the QNMs of a test scalar field in the background of the TS varying the ratio $r_B/r_S$ for different values of the angular momentum number $l=1,2,3$. Note that these modes have been already studied in detail in the literature, see~\cite{Heidmann:2023ojf,Bianchi:2023sfs}. 
The QNMs are computed by solving, with a direct integration method~\cite{Pani:2013pma,Pani:2012bp}, the eigenvalue problem represented by the homogeneous version of Eq.~\eqref{eq:KG_eq_radial} with $n=0$, together with proper boundary conditions (see~\cite{Dima:2024cok,Dima:2025zot} for more details).

\begin{figure}[th]
  \centering
\includegraphics[width=0.485\textwidth]{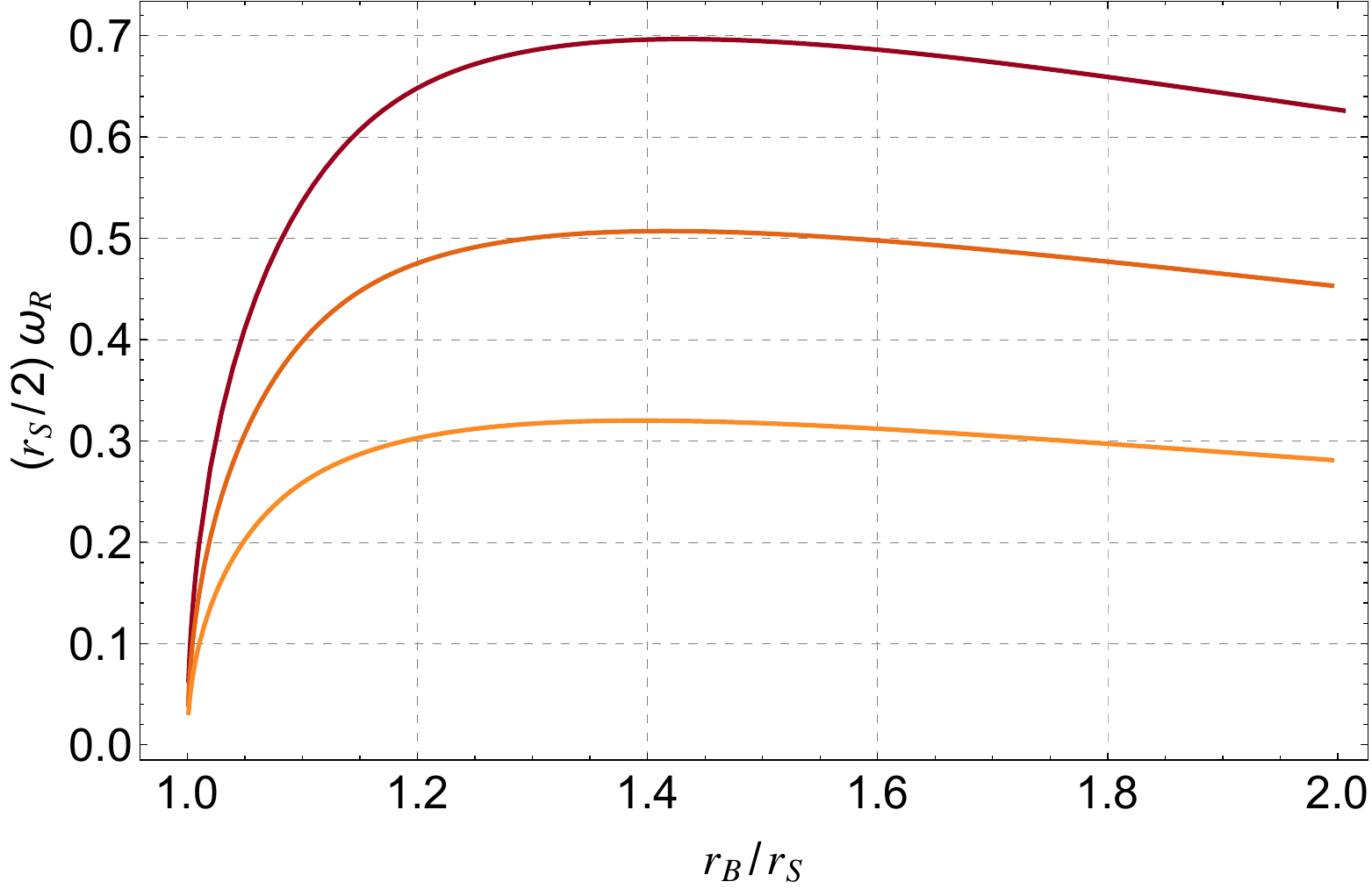}\\
\includegraphics[width=0.485\textwidth]{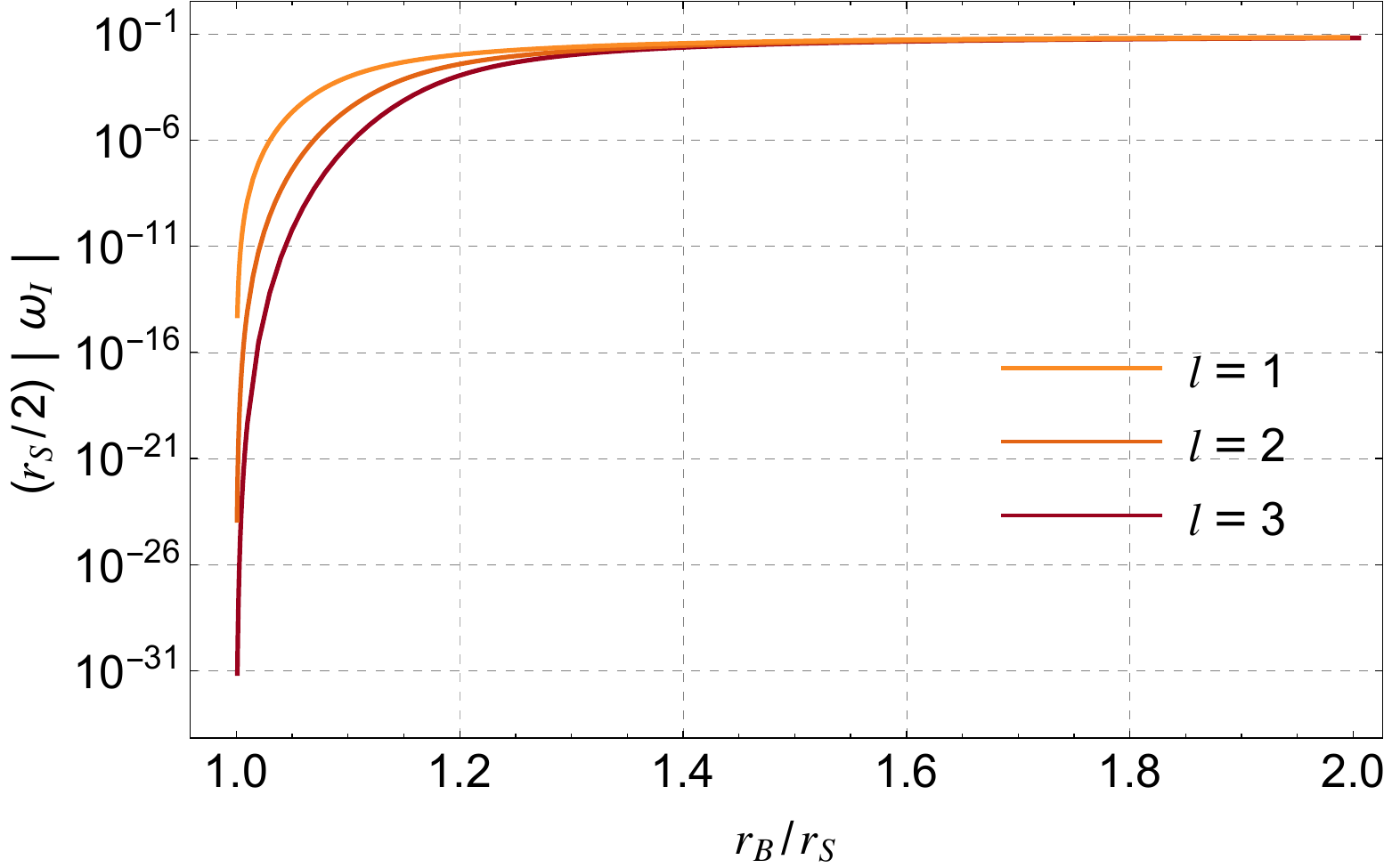}
  \caption{Real part (top panel) and imaginary part (bottom panel) of the fundamental QNMs frequencies for test scalar perturbations in the background of the TS varying the ratio $r_B / r_S$ for different values of $l = 1, 2, 3$.
  }\label{fig:QNMs_TS_rS}
\end{figure}

As shown in Fig.~\ref{fig:QNMs_TS_rS}, as the $r_B/r_S=1$ limit is approached, the fundamental mode becomes increasingly long-lived, and its real part tends to zero, thus making the scalar energy flux at infinity more prone to develop a resonance. As the angular momentum number $l$ increases, the absolute value of the imaginary part of the frequencies significantly decreases \cite{Bianchi:2023sfs}, making the resonance of the energy flux extremely narrow. 

\bibliographystyle{apsrev4-1}
\bibliography{top_star.bib}

\end{document}